%% file: ISSA.tex
\documentclass{llncs}[10]
\usepackage{amsmath}
\usepackage{amsfonts}
\usepackage{amssymb}
\usepackage{ifpdf}
\usepackage{subfig}
\usepackage{wrapfig}
\usepackage{cite}

\ifpdf

  \usepackage[pdftex]{epsfig}
  \usepackage[pdftex]{hyperref}

\else

    \usepackage[dvips]{epsfig}
    \newcommand{\href}[2]{#2}

\fi

\begin{document}

\title{Identifying Shapes Using Self-Assembly (extended abstract)}%
\author{Matthew J. Patitz\thanks{Department of Computer
Science, University of Texas--Pan American, Edinburg, TX, 78539, USA.
mpatitz@cs.panam.edu.} \and Scott M. Summers\thanks{Department
of Computer Science and Software Engineering, University of Wisconsin--Platteville, Platteville, WI 53818, USA. summerss@uwplatt.edu.}}

\institute{}

\date{}

\maketitle

%\title{Identifying Shapes Using Self-Assembly (extended abstract)\footnote{This research was supported in part by National
%   Science Foundation Grants 0652569 and 0728806. }}%
%\author{Matthew J. Patitz\footnote{ Department of Computer
%Science, University of Texas--Pan American, Edinburg, TX, 78539, USA.
%mpatitz@cs.panam.edu. } \and Scott M. Summers.\footnote{Department
%of Computer Science, Iowa State University, Ames, IA 50011, USA.
%summers@cs.iastate.edu. This author's research was supported in part
%by NSF-IGERT Training Project in Computational Molecular Biology
%Grant number DGE-0504304}}
%\date{}
%\maketitle

% ----------------------------------------------------------------
\begin{abstract}
In this paper, we introduce the following problem in the theory of algorithmic self-assembly: given an input shape as the seed of a tile-based self-assembly system, design a finite tile set that can, in some sense, uniquely identify whether or not the given input shape--drawn from a very general class of shapes--matches a particular target shape. We first study the complexity of correctly identifying squares. Then we investigate the complexity associated with the identification of a considerably more general class of non-square, hole-free shapes.
\end{abstract}

\input{introduction}
\input{prelims}
\input{planar_squares}
\input{non_planar_more_shapes}
\input{non_planar_more_shapes_two_stages}

\input{conclusion}

\subsubsection*{Acknowledgments}Both authors would like to thank Paul Rothemund and John Mayfield for helpful discussions and useful feedback.

%----------------------------------------------------------------
\bibliographystyle{amsplain}
\bibliography{main,tam,otherSA}

\clearpage

\input{appendix}

\end{document}

%% file: introduction.tex
\section{Introduction}

As amazingly complex as biological organisms are, at the nanoscale they are composed of ``simple'' pieces that spontaneously self-assemble--a bottom-up process by which a relatively small group of fundamental components combine according to local rules in order to form a complex structure. This very basic process is responsible for the vast diversity and complexity of life--from the most simple single-cell organisms to human beings.

Inspired by nature, scientists have developed and studied a wide variety of artificial self-assembling systems in order to produce structures as varied as nanowires \cite{SANanowires}, crystals \cite{SACrystals}, nanofiber scaffoldings \cite{SANanofiberScaffold}, landscapes for nanoscale robots \cite{WinfreeDNARobots2010,SeemanDNARobots2010} and dozens of other novel supramolecules (see \cite{RotOrigami06,OrigamiBox,SANanocrystalSheets} for more examples).  In addition to experimental work, there has also been a plethora of theoretical work in the design and analysis of the complexities and limitations of self-assembling systems, with notable examples including \cite{RotWin00,AGKS05,FuSch09,MLR07}.

Much of the research in algorithmic self-assembly (both theoretical and experimental) can be loosely categorized into four ``genres:'' the self-assembly of shapes \cite{RoPaWi04,SolWin07}, evaluating computable functions to direct nanoscale self-assembly \cite{Winfree96universalcomputation,CCSA}, replicating input shapes \cite{Replication10}, and creating novel materials that have various chemical properties \cite{zeng2002exc}. In this paper, we introduce a novel (theoretical) self-assembly problem that is motivated by not only the behavior of biological systems but also the practical need to verify artificial laboratory-based self-assembly systems. We call this new problem the \emph{shape identification problem}, and define it as the task of designing a tile-based self-assembly system that positively identifies a target structure that has a pre-specified shape (and size) from among possible ``junk'' structures drawn from a very general pool of objects.

\textbf{Motivation:} Shape identification is a fundamental process of nature and is explicitly used by biological systems in a variety of ways. First and foremost, the immune system generates complexes whose express purpose is to selectively identify--and ultimately bind to--precisely-shaped locations on the surface of foreign objects in order to mark them for destruction (by, for example, killer T cells).  Also, cellular transport systems, such as those which transport amino acids or sugars, work by moving specifically-shaped molecules from one side of a membrane to the other. Furthermore, the power of a self-assembling system (natural or artificial) ultimately arises from the information encoded in its constituent components. In the notable case of proteins, it is the information embedded in their precise three-dimensional geometry that allow them to match and combine with the necessary specificity to build the fundamental building blocks of life.

The ability to correctly identify only the completely formed products of an artificial self-assembly system is also of extreme importance to practitioners. Unfortunately, accomplishing this task is difficult because the self-assembly environment is often variable and chaotic, where mistakes are likely to be made and partially-formed products common. Current methods of imaging the results of nanoscale self-assembling systems provide insufficient resolution for automated visual inspection of assemblies and require error-prone manual inspection (for instance, by pouring over atomic force microscope images). Methods such as gel electrophoresis allow for the separation of products based loosely on their mass and shape, but unfortunately with far less shape specificity than desired. With accurate nanoscale shape identification schemes, however, the accuracy of the techniques that experimenters use to identify the products of self-assembling systems could be improved dramatically.

In this paper, we formulate the shape identification problem in algorithmic self-assembly (defined formally in Section~\ref{subsection_formulation}) and exhibit a variety of solutions thereof while working in the \emph{RNAse enzyme model}--a discrete mathematical model of two-handed tile-based self-assembly (based on Winfree's abstract Tile Assembly Model \cite{Winf98,RotWin00}) that distinguishes DNA tiles from RNA tiles and permits the usage of an RNAse enzyme that dissolves all of the RNA tiles in a given assembly. This model was initially suggested by Rothemund and Winfree in the final section of \cite{RotWin00} and formally defined by Abel, Benbernou, Damian, Demaine, Demaine, Flatland, Kominers and Schweller \cite{Replication10}. We focus our attention on the design of ``small'' tile sets that identify certain types of target shapes by tagging them with a border of DNA tiles. Note that the borders, which signify positive identification, could also be ``functionalized'' with bindings sites that facilitate the easy extraction of only the correct assemblies. It is worthy of note that, while the results presented in this paper are based on tile-based self-assembly systems identifying tile-based assemblies, the underlying principles of this paper are applicable to the identification of any type of precisely shaped shaped object (e.g., a DNA origami complex \cite{RotOrigami06}) so long as its perimeter advertises the necessary binding domains, which in the case of this paper, are single-stranded DNA sequences.

\textbf{Statement of Results:} In Section~\ref{planar_square}, we exhibit a \emph{planar} tile assembly system (a.k.a., a system in which all supertiles have obstacle-free paths to their mates and therefore require the use of only two spacial dimensions; see \cite{DDFIRSS07} for more additional examples of planar self-assembly systems) capable of identifying any $n \times n$ square using $O(\log n)$ unique tile types. We then use a well-known optimal encoding scheme \cite{ACGHKMR02,SolWin07} to reduce the number of unique tile types in the aforementioned construction (while preserving planarity) to $O\left(\frac{\log n}{\log \log n}\right)$. We subsequently prove a matching lower bound on the minimum number of unique tile types required to identify an $n \times n$ square. This implies that the complexity of identifying an $n \times n$ square in a restricted RNAse enzyme model coincides exactly with that of its self-assembly in the abstract Tile Assembly Model \cite{ACGHKMR02}. We conclude Section~\ref{planar_square} with a $O(1)$ size planar tile assembly system that ``universally'' identifies whether or not any hole-free input shape is an $n \times n$ square. In Section \ref{non_planar_more_shapes}, we develop a non-planar tile assembly system that identifies a wide variety of ``hole-free'' shapes that have a kind of ``perimeter-rectangle decomposition'' that uses an optimal number of unique tile types in the sense of Kolmogorov complexity. We then mildly extend the aforementioned result to identify a more general class of shapes--assuming the use of two different types of RNAse enzymes is permitted. 

%% file: prelims.tex
\section{Preliminaries and Notation}
\label{prelim}

Please see Section \ref{two-handed} for a brief description of the two-handed Tile Assembly Model and Section \ref{subsection_rna_model} for a description of the RNAse enzyme extension to it.

\subsection{Formulation of the Shape Identification Problem in the RNAse Enzyme Model}
\label{subsection_formulation}
Fix a temperature $\tau \in \mathbb{N}$. For every \emph{shape} (a.k.a., a finite, connected subset of $\mathbb{Z}^2$) $X$, define the assembly $\sigma_X$ as the placement of specially designated seed (DNA) tiles at every point in $X$ subject to the restrictions that $\sigma_X$ must be $\tau$-stable, the strengths of all of the ``external'' glues must be $1$ and there should be no way to determine ``corner tiles'' of $\sigma_X$ (this latter restriction is accomplished by assuming that all external glues of $\sigma_X$ are labeled with the empty string). \footnote{We speculate that one possible molecular implementation of this might be achieved using Rothemund's DNA origami as a seed structure \cite{RotOrigami06,BarSchRotWin09} to which DNA and RNA tiles can subsequently attach.}

We are now ready to define the \emph{shape identification problem} in self-assembly. Fix some class of shapes $\mathcal{C}$ along with a target shape $X \in \mathcal{C}$. The goal is to design a finite set of tile types $T_X$ (that does not contain any of the tile types that appear in the seed assembly $\sigma_X$) satisfying the following condition: given an input shape $Y \in \mathcal{C}$ encoded as the seed assembly $\sigma_Y$, if $X = Y$ then the system $\left( T_X, \sigma_Y, \tau\right)$ uniquely produces a fully-connected final assembly $\alpha$ consisting of $\sigma_Y$ with a fully connected ring of ``border'' tiles along the border of $\sigma_Y$; if $X \ne Y$, then $\sigma_Y$ is the uniquely produced terminal structure. If it is possible to accomplish such a task (i.e., design such a tile set $T_X)$, then we say that $T_X$ \emph{identifies} the shape $X$ \emph{with respect to} $\mathcal{C}$. Note that since we must add the RNAse enzyme last, the border tiles must necessarily be DNA tiles so that they are not simply dissolved away at the end. We say that a shape $X$ can be \emph{identified with respect to} $\mathcal{C}$ if there exists a finite set of tile types $T_X$ that can identify it with respect to $\mathcal{C}$.
\begin{figure}[htp]
%previous widths:  1.25, 1.78, 2.09
\centering
    \subfloat[][Target shape $X$ and input shape $Y$]{%
        \label{fig:inputshape}%
        \includegraphics[width=1.15in]{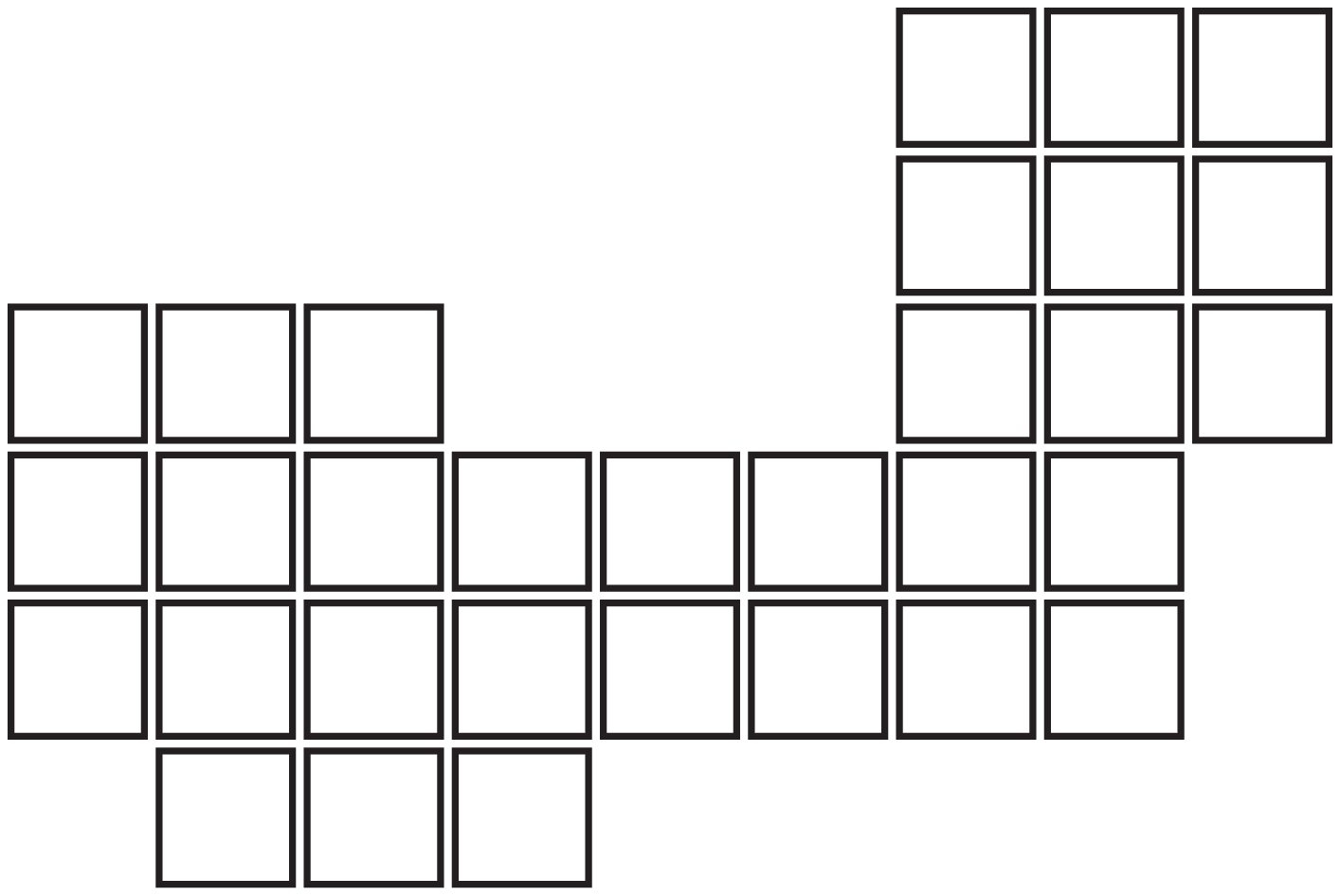}}%
        \hspace{20pt}%
    \subfloat[][Seed assembly $\sigma_Y$]{%
        \label{fig:inputassembly}%
        \includegraphics[width=1.3in]{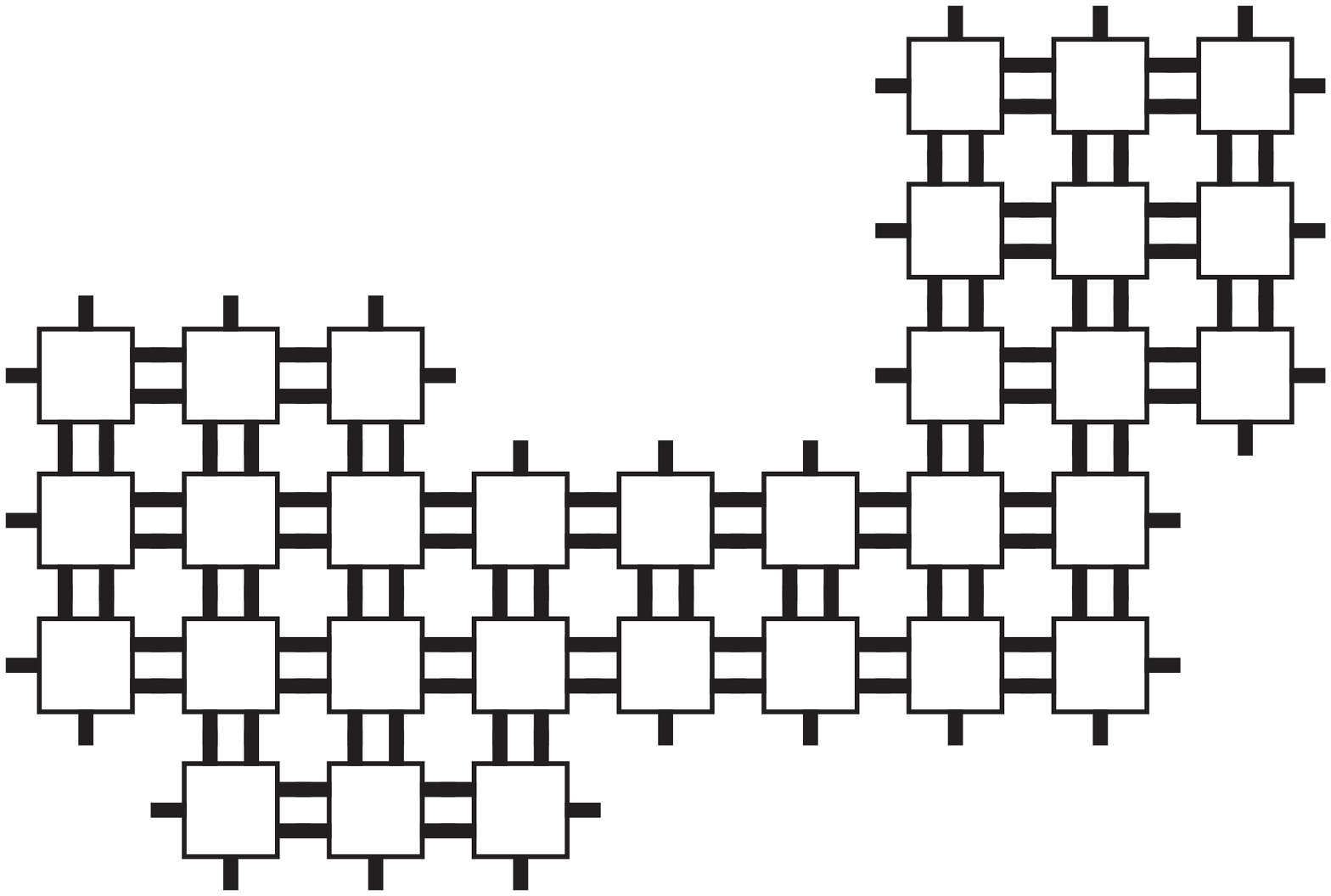}}
        \hspace{20pt}%
     \subfloat[][The goal!]{%
        \label{fig:finalassembly}%
        \includegraphics[width=1.50in]{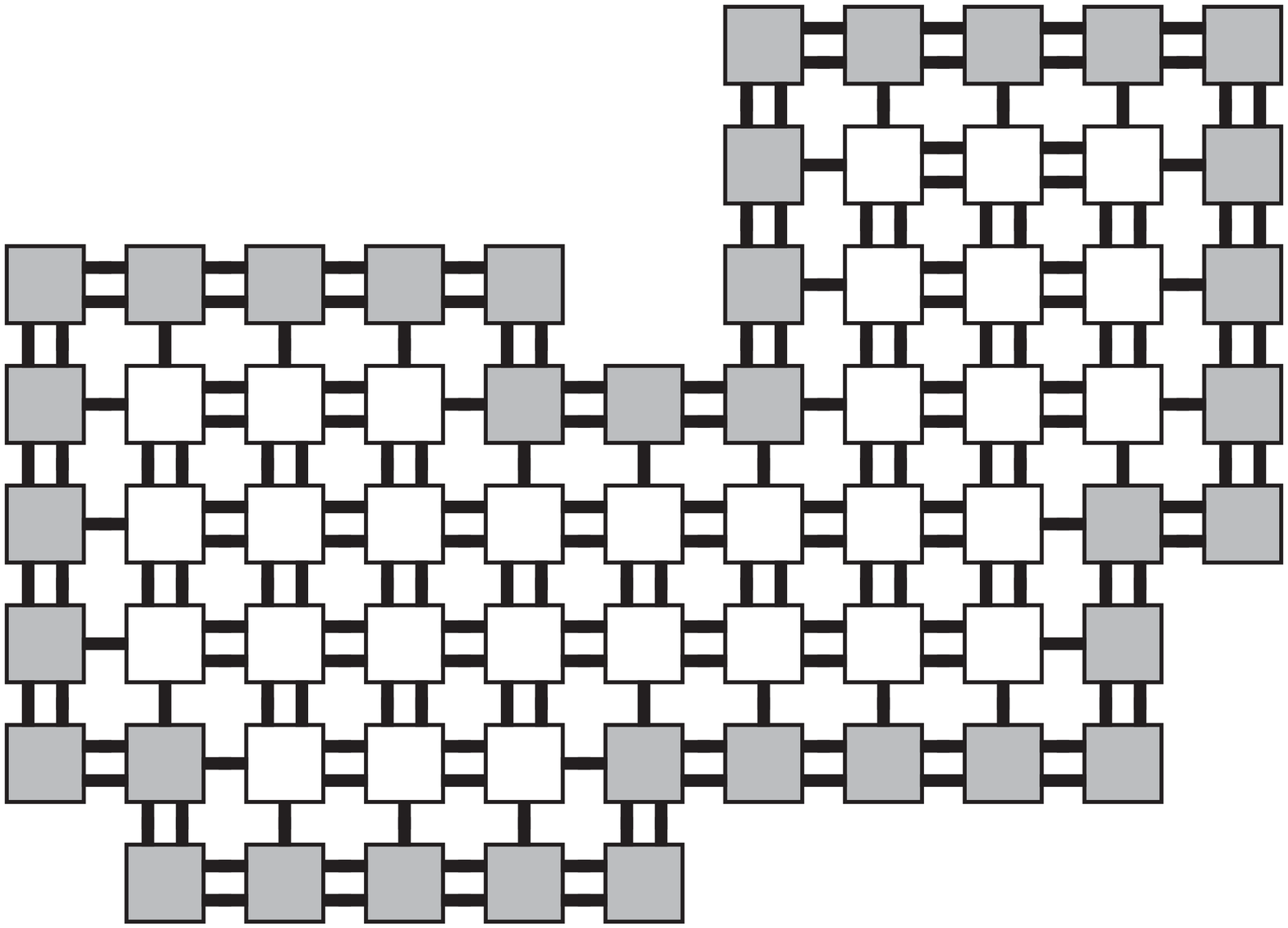}}%\\
    \caption{\small The desired outcome for a ``yes'' instance of the shape identification problem.}%
        \label{fig:shapeidenfiticationexample}%
\end{figure}

An example of an instance of the shape identification problem (for some shape with respect to some class of shapes) is depicted in Figure~\ref{fig:shapeidenfiticationexample}. We say that the \emph{identification complexity of} $X \in \mathcal{C}$ \emph{with respect to} $\mathcal{C}$ is the minimum number of tile types necessary to identify it with respect to $\mathcal{C}$ (this is analogous to the \emph{tile complexity} of a shape $X$ being defined as the minimum number of tile types necessary to uniquely produce $X$).

%% file: planar_squares.tex
\section{Identification of $n \times n$ Squares with Planarity}
\label{planar_square}

In this section, we exhibit two \emph{planar} self-assembly systems (a.k.a., systems in which all supertiles have obstacle-free paths to their mates and therefore require use of only two spacial dimensions; see \cite{DDFIRSS07} for more discussion of planarity) that efficiently identify $n \times n$ squares with respect to the set of all \emph{hole-free} shapes: shapes whose complements are infinite, connected subsets of $\mathbb{Z}^2$. We also construct a universal tile set that is capable of identifying whether a given input shape is in fact a square of \emph{any} dimension.

For each $n \in \mathbb{N}$, let $S_n = \{0,1,\ldots,n-1\}^2$ be the $n \times n$ square whose lower-left corner is positioned at the origin. Throughout this paper, $\mathcal{C}$ denotes the class of all hole-free shapes. The motivating factor behind defining $\mathcal{C}$ this way is because we want our constructions to be able to distinguish a target shape from among many different possible ``junk'' (i.e., non-square) input shapes. The temperature for all of our constructions in this paper is $\tau = 4$.

\subsection{Planar Identification of $n \times n$ Squares with $O(\log n)$ Unique Tile Types}
Our first main result of this section is the following theorem, which states that there is an efficient planar identification scheme for $n \times n$ squares.

\begin{theorem}
\label{theorem_square_log_n}
For all $6 < n \in \mathbb{N}$, the identification complexity of $S_n$ with respect to $\mathcal{C}$ is $O(\log n)$.
\end{theorem}

The proof idea of Theorem~\ref{theorem_square_log_n} is as follows. Suppose we are trying to identify $S_n$ for some $6 < n \in \mathbb{N}$. Given an input shape $Y \in \mathcal{C}$, our construction first attaches ``verification modules'' to north- south- and west-facing sides of $Y$ (if $Y$ is an $n \times n$ square, then there will be exactly one of each of these types of sides). These modules are side-by-side pairs of unary counters and binary counters that do not interact with each other as they count. The unary counters count (in unary) the length of the side to which they are attached and the binary counters essentially count (in binary) up to $n$ (the dimension of the target square). Each verification module compares the length of the side to which it is attached with $n$. If all three verification modules report success \emph{and} agree with each other, then the input shape is in some sense ``almost'' a square. The three verification modules then cooperate to allow DNA border tiles to start attaching to the east-facing side of the input shape. If border tiles can attach to all but the two bottom rightmost points along the east-facing side, then the input shape is in fact $S_n$ and our construction reaches an intermediate terminal state. At this point, we add the RNAse enzyme leaving only the input shape to which the east-facing border tiles are attached. The remaining border tiles attach in a clockwise fashion until a complete and fully connected border is assembled. However, if not all east-facing border tiles can attach (i.e., the input shape is not $S_n$), then after the RNAse enzyme is added all previously-attached border tiles will disassociate one at a time until no tiles are attached to the input shape.

Please see Section \ref{planar_construction_sketch} for a more detailed explanation of this construction.

\begin{figure}[htp]
\centering
    \includegraphics[width=\textwidth]{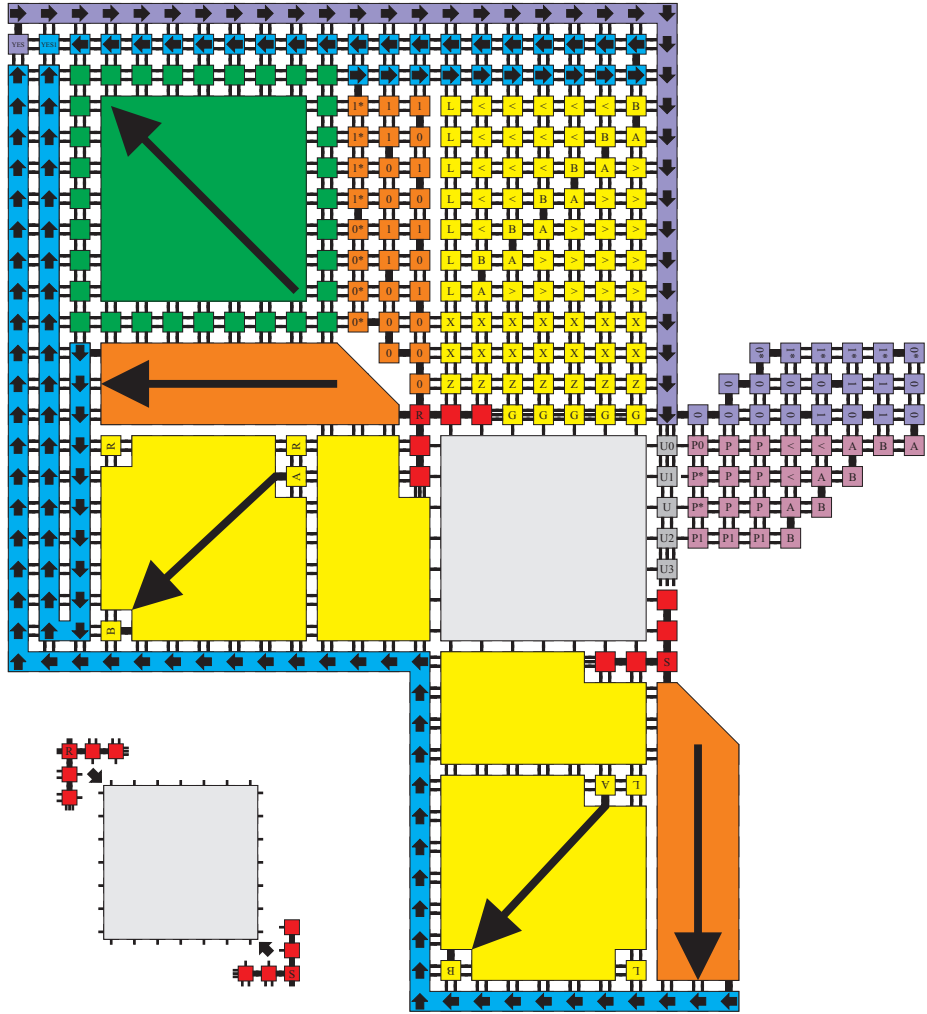}
\caption{\label{fig:log_construction_overview} An example of our construction for Theorem~\ref{theorem_square_log_n} with $n = 7$. Our tile set is partitioned into several logical groups--each given a different color in this figure to represent the relative order in which they assemble (i.e., Red, Orange, Yellow, Green, Blue, Indigo, Violet). First, the red supertiles assemble and attach to the corners of the input shape. The orange group essentially encodes the length of the to-be-identified square via a binary counter and requires $O(\log n)$ unique tile types. The ``U'' border tiles attach along the east-facing side of the input shape. All tiles are RNA tiles except for the ``U'' tiles and, of course, the tiles that make up the initial seed square. }
\end{figure}

\subsection{$\Omega\left(\frac{\log n}{\log\log n}\right)$ Unique Tile Types are Necessary to Identify an $n \times n$ Square}
In \cite{RotWin00}, Rothemund and Winfree established an $\Omega\left( \frac{\log n}{\log \log n}\right)$ lower bound on the number of tile types required to uniquely assemble an $n \times n$ full square for almost all $n$. In this section, we adapt their information-theoretic proof technique to the shape identification problem for $n \times n$ squares under the RNAse enzyme model.

\begin{theorem}
\label{theorem_square_lower_bound}
For all but finitely many $n \in \mathbb{N}$, the identification complexity of $S_n$  with respect to $\mathcal{C}$ is $\Omega\left(\frac{\log n}{\log\log n}\right)$.
\end{theorem}
\begin{proof}[sketch]
For each $n \in \mathbb{N}$, define the \emph{Kolmogorov complexity of} $n$ as $K(n) = \min\{ |\pi| \mid U(\pi) = n \}$ where $U$ is some fixed universal Turing machine. The reader is encouraged to consult \cite{li_vitanyi_1997} for a more detailed discussion of Kolmogorov complexity. An easy application of the pigeonhole principle tells us that for almost all $n \in \mathbb{N}$, $K(n) = \Omega(\log n)$.

Note that for a given $n \in \mathbb{N}$ and temperature $\tau \in \mathbb{N}$, there exists a constant size Turing machine $M$ that takes as input a tile set $T_{n}$ that uniquely identifies $S_n$, a seed assembly representing the input shape $\sigma$ (as discussed in Section~\ref{subsection_formulation}) and outputs the maximum extent (height or width) of the uniquely produced terminal assembly. We can then use $M$ as a subroutine in another constant size Turing machine $N$ that takes as input $T_{n}$ and sequentially simulates $M$ on $T_{n}$ with the seed assembly $\sigma_{S_i}$ for $i \geq 0$ \emph{in order} while checking if the maximum extent (height or width) of the $i^{\textmd{th}}$ uniquely produced terminal assembly is $i+2$. Since $T_{n}$ uniquely identifies $S_n$, we are guaranteed that this search will eventually terminate, at which point $N$ halts and outputs $i = n$. This implies that the size of (number of bits in) the encoding for $T_n$ must be $\Omega(\log n)$. Since we can encode an arbitrary tile set $T$ with $O(|T|\log |T|)$ bits (assuming $T$ has a diagonal strength function), we have that $|T_n| = \Omega\left(\frac{\log n}{\log \log n}\right)$.
\end{proof}

\subsection{Planar Identification of $n \times n$ Squares with $O\left(\frac{\log n}{\log\log n}\right)$ Unique Tile Types}
\label{section_squares_optimal}
The construction for Theorem~\ref{theorem_square_log_n} can be modified to prove the following asymptotically optimal result for the identification of $n \times n$ squares.

\begin{theorem}
\label{theorem_square_optimal}
For all $6 < n \in \mathbb{N}$, the identification complexity of $S_n$ with respect to $\mathcal{C}$ is $O\left(\frac{\log n}{\log\log n}\right)$.
\end{theorem}

Consult Section~\ref{square_optimal_proof_sketch} for more details.

\subsection{Universal Planar Identification of Squares with $O(1)$ Unique Tile Types}
In the previous subsections, we focused our attention on the problem of identifying $n \times n$ squares for particular values of $n$ from among any input shape drawn from the set of all hole-free shapes. We now study the related problem of universally identifying whether or not a given input shape is an $n \times n$ square for some $n \in \mathbb{N}$. Here, we are given an arbitrary hole-free input shape and we wish to correctly identify it (in the sense of tagging its border with special tiles) if and only if it is in fact a square.

\begin{theorem}
\label{universal_theorem} There exists a planar (universal) tile set $T$ with $|T| = O(1)$ such that, for all $6 < n \in \mathbb{N}$, $T$ identifies $S_n$ with respect to $\mathcal{C}$.
\end{theorem}

Intuitively, we prove Theorem~\ref{universal_theorem} by constructing a constant size tile set that (1) grows unary counters off of the north, west and south sides of the input shape and then (2) allows a border of DNA tiles to assemble if and only if all of the counters agree on the same value (in addition to the right side of the input shape being consistent with that of a square).  Please see Section \ref{universal_squares_proof_sketch} for a more detailed discussion.

Note that planarity isn't necessary for any results presented in this section.

%% file: non_planar_more_shapes.tex
\section{Non-Planar Identification of More Shapes}
\label{non_planar_more_shapes}

We now exhibit a non-planar self-assembly system that efficiently identifies a wide variety of shapes with respect to the set of all hole-free shapes but at the expense of sacrificing planarity.  We first define some notation.

\begin{figure}[htp]
\centering
    \subfloat[][An example shape $X$]{%
        \label{fig:square_decomp_example1}%
        \includegraphics[width=1.5in]{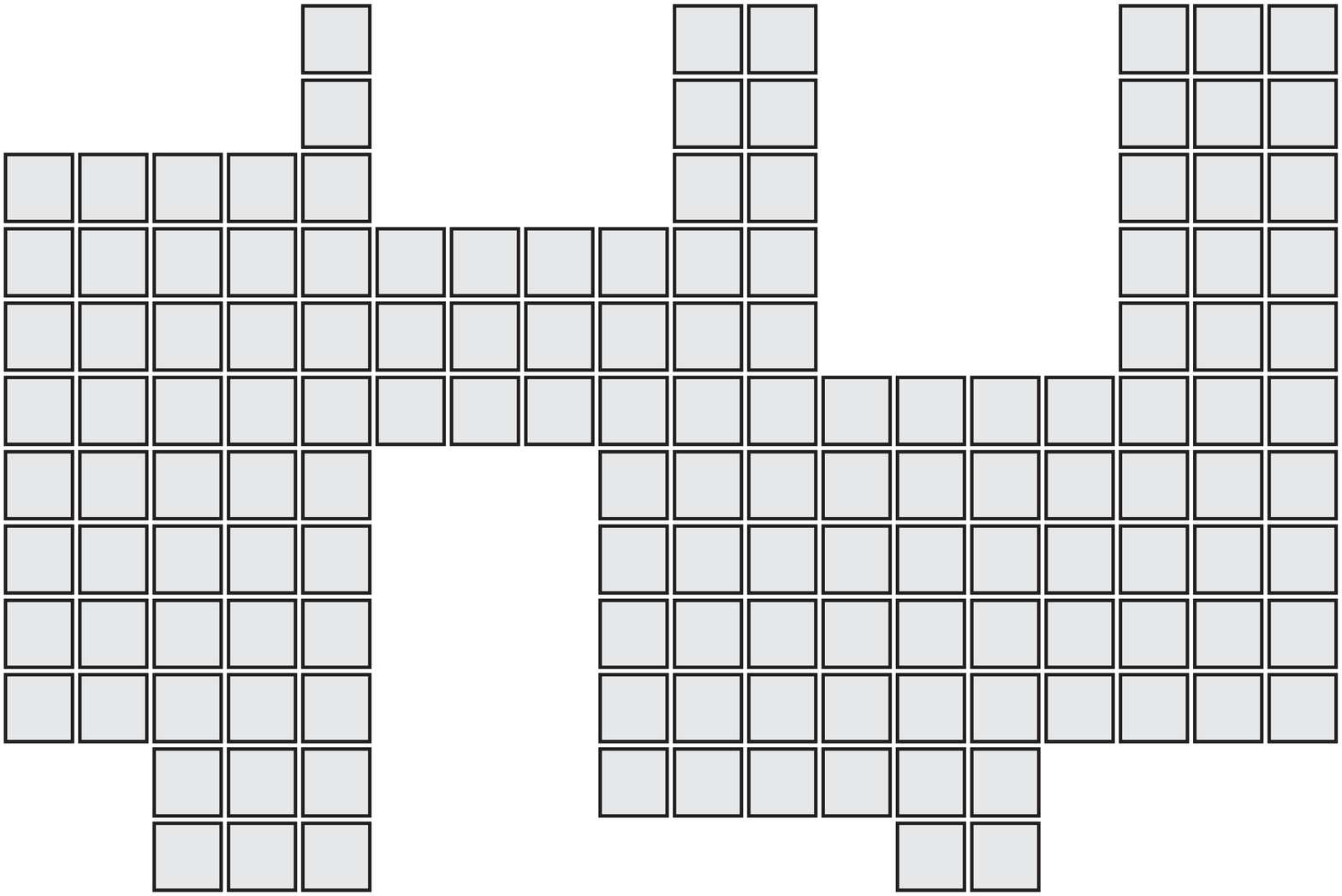}}%
        \hspace{40pt}%
     \subfloat[][A valid perimeter-rectangle decomposition of $X$.]{%
        \label{fig:square_decomp_example2}%
        \includegraphics[width=1.5in]{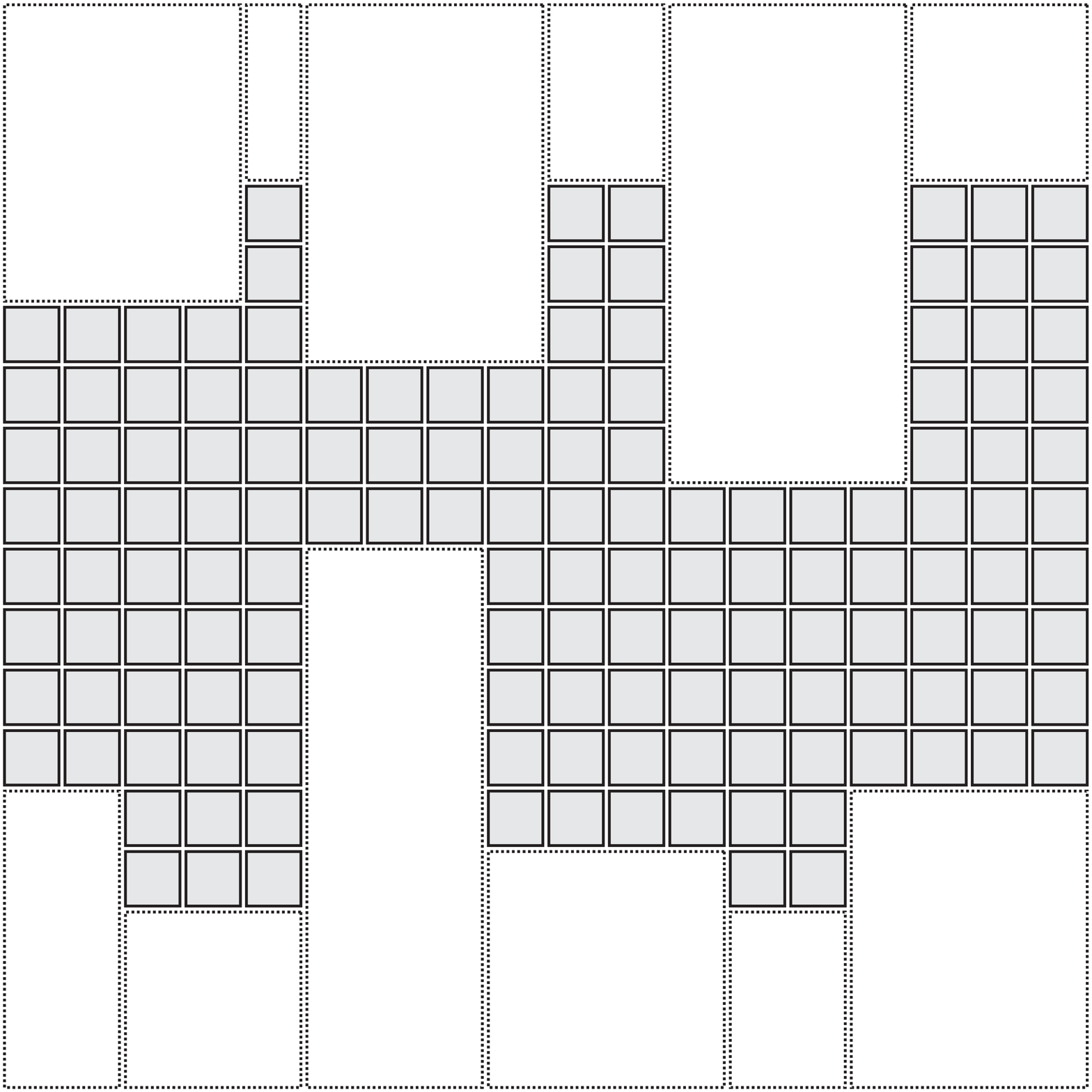}}%\\
    \caption{\label{fig:square_decomp_example} An example of a perimeter-rectangle decomposition of a particular shape.}%
\end{figure}

Let $ (x,y) = \vec{a} \in \mathbb{Z}^2$ and $(w,z) = \vec{b} \in \mathbb{Z}^2$ and define $d_{\infty}\left(\vec{a},\vec{b}\right) = \max\{ |x-w|,|y-z|\}$.  If $X$ is a shape, then we say that the \emph{feature size} of $X$ is the minimum $d_{\infty}\left(\vec{a},\vec{b}\right)$ such that $\vec{a}$ and $\vec{b}$ are on two non-adjacent edges of $X$. We say that a shape $X$ is $x$-\emph{monotone} if its intersection with any vertical line is a connected line. If $X$ is a shape, then let $\widetilde{R}(X)$ be the smallest rectangle that contains $X + \{(0,3),(0,-3)\}$, where, for any set $A \subseteq \mathbb{Z}^2$, $X + A = \left\{ \vec{x} + \vec{a} \; \left| \; \vec{x}\in X \textmd{ and } \vec{a} \in A\right.\right\}$. We say that $X$ has a \emph{perimeter}-\emph{rectangle decomposition}, denoted as $\{R_i\}_{i=0}^{n-1}$ for some $n \in \mathbb{N}$, if: for each $0 \leq i < n$, $R_i$ is a rectangle, for all $0 \leq j < n$, $i \ne j \Rightarrow R_i \cap R_j = \emptyset$,  $height(R_i) \leq 2^{width(R_i)} + 3$, $\widetilde{R}(X) - X = \bigcup_{i=0}^{n-1}{R_i}$, and for each $0 \leq i < n$, the perimeter of $R_i$ intersects the perimeter of $\widetilde{R}(X)$. For any rectangle $R$, we write $h(R) = height(R)$, and $w(R) = width(R)$. See Figure~\ref{fig:square_decomp_example} for an example of a shape and a valid perimeter-rectangle decomposition thereof. Recall that $\mathcal{C}$ is the set of all hole-free shapes.

\begin{theorem}
\label{non_planar_technical_theorem}
Fix a universal Turing machine $U$. Let $X$ be a shape and $\pi_X$ be any program such that $U(\pi_X) = \langle X \rangle$, where $\langle \cdot \rangle$ is a standard binary encoding of a finite object. If $X$ is $x$-monotone, has feature size $5$ and has perimeter-rectangle decomposition $\{R_i\}_{i=0}^{n-1}$, then the identification complexity of $X$ with respect to $\mathcal{C}$ is $O\left(\frac{\left|\pi_X\right|}{\log \left| \pi_X \right|}\right)$.
\end{theorem}

Note that by choosing $\pi_X$ to be the shortest program such that $U(\pi_X) = \langle X \rangle$, then $|\pi_X| = K(X)$.  The proof idea of Theorem~\ref{non_planar_technical_theorem} is as follows. Given a shape $X$ that satisfies the hypothesis, our construction first converts $X$ into a string $xyz$ such that $x$ encodes all of the ``north-facing'' features of $X$, $y$ encodes $h\left(\widetilde{R}(X)\right)$ and $z$ encodes all of the ``south-facing'' features of $X$. Our construction then uses this string as a seed in order to assemble a frame to which an input shape can attach. Once the input shape attaches to the frame, a single-tile-wide border assembles around the perimeter of the input shape and fills in completely if and only if the input shape matches the target shape. Once the RNAse enzyme is added, the frame dissolves and if the input shape has a full border of DNA tiles, then we are done and the target shape has been correctly identified. However, if the input shape does not match the target shape, then our construction ensures that a full border around the input shape is not allowed to assemble. Moreover, if the border is not fully formed after the RNAse enzyme is added, then the partially formed border will disassemble in a counter-clockwise fashion one tile at a time eventually leaving the input shape completely free of DNA border tiles. We encode $X$ as a program $\pi_X$ using the optimal encoding scheme of Soloveichik and Winfree \cite{SolWin07}, whence the identification complexity of $X$ with respect to $\mathcal{C}$ is $O\left( \frac{\left| \pi_X \right|}{\log \left| \pi_X \right|} \right)$.  Please see Section \ref{non-planar_proof_sketch} for more details of this construction.

%% file: non_planar_more_shapes_two_stages.tex
\section{Non-Planar Identification of \emph{Even More} Shapes}
\label{non_planar_more_shapes_two_stages}
\begin{figure}[htp]
\centering
    \subfloat[][Very high-level overview of the construction for Theorem~\ref{non_planar_two_dissolve_theorem}. The grey wedge represents a self-assembly simulation of a Turing machine that unpacks a compact description of all of the rectangles that eventually assemble into a frame that accepts the input shape.]{
    \label{fig:non_planar_two_dissolve_overview_1}
    \includegraphics[width=2.5in]{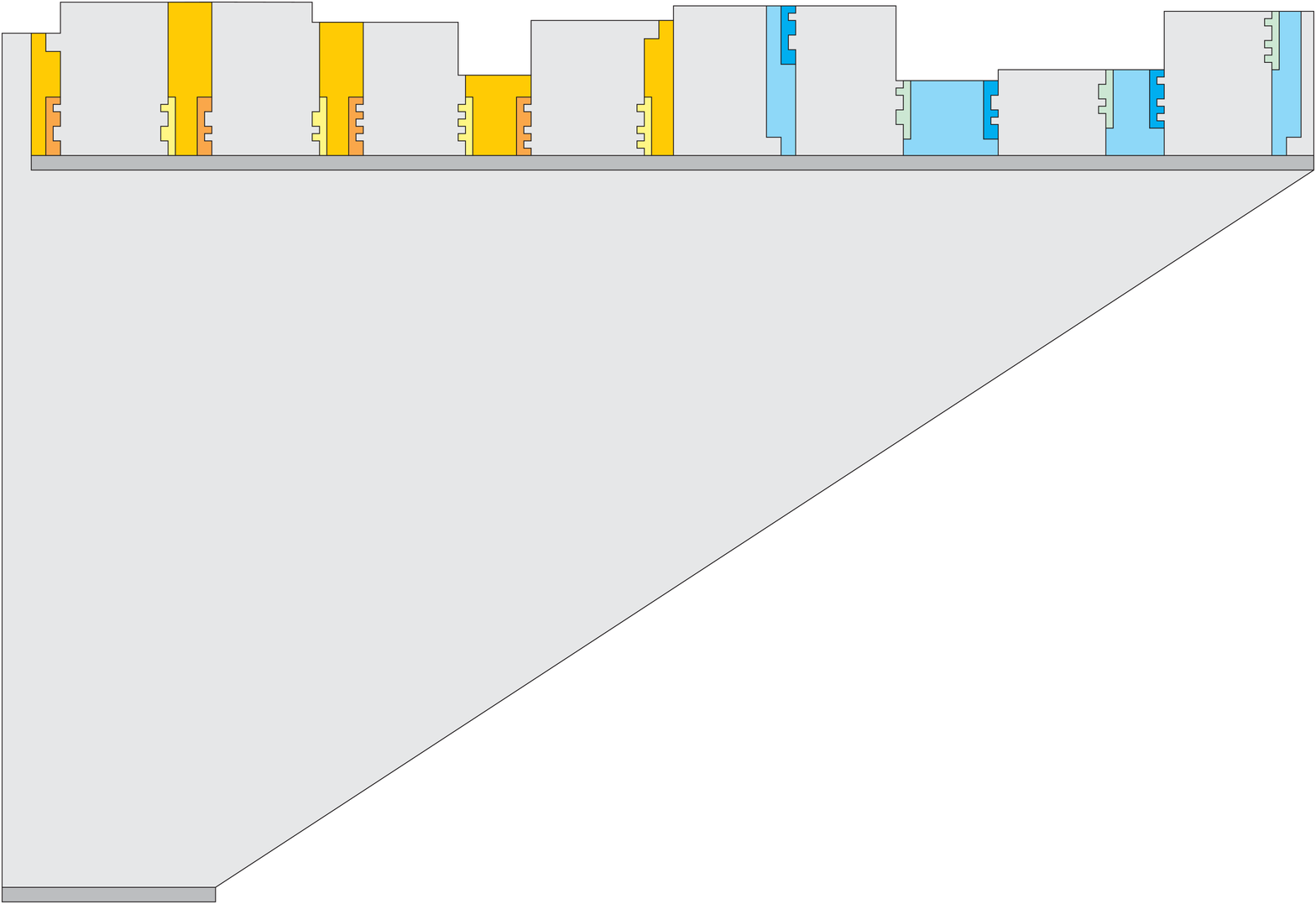}}
    \hspace{20pt}
    \subfloat[][After the first type of RNAse is added, all of the supertiles are free to assemble into a frame that accepts the input shape.]{%
        \label{fig:non_planar_two_dissolve_final}%
        \includegraphics[width=1.3in]{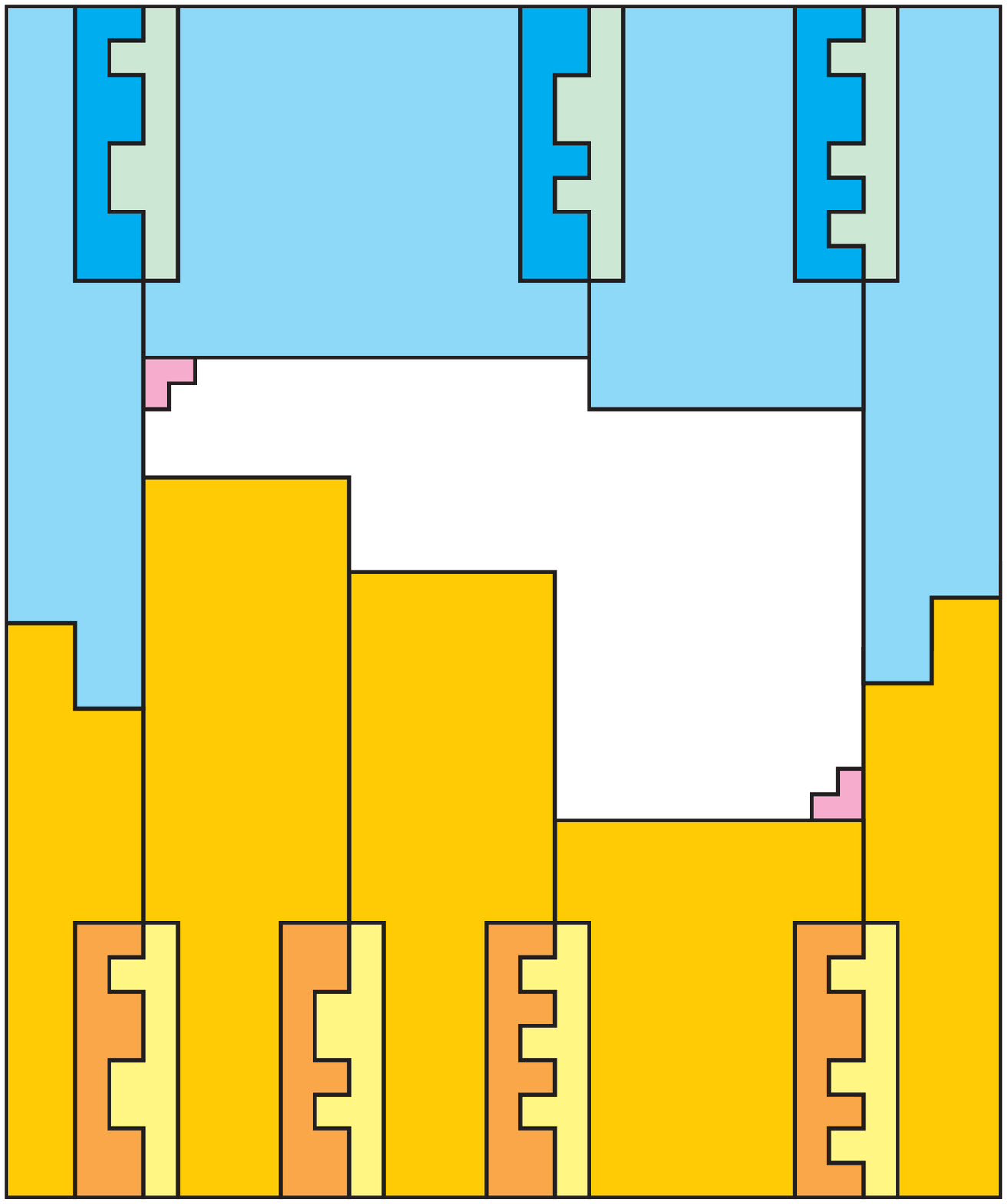}}
    \caption{\label{fig:non_planar_two_overview} Overview of the construction for Theorem~\ref{non_planar_two_dissolve_theorem}.}
\end{figure}
Throughout this paper, we have assumed that the RNAse enzyme (the agent responsible for removing all of the RNA tile types) is added once--and only--after the initial stage of two-handed self-assembly is allowed to reach a terminal state. Under this assumption, the RNAse enzyme universally dissolves \emph{every} RNA tile in all of the produced assemblies. In this section, we relax this restriction and allow for the use of \emph{two} different types of RNAse enzymes in two separate dissolve stages that each affect a different group of RNA tiles. Doing so leads to a mild refinement of Theorem~\ref{non_planar_more_shapes}, stated precisely as the following theorem.

\begin{theorem}
\label{non_planar_two_dissolve_theorem}
Fix a universal Turing machine $U$ and let $X$ be a shape and $\pi_X$ be any program such that $U(\pi_X) = \langle X \rangle$. If $X$ is $x$-monotone, has feature size $6$ and if the use of two different types of RNAse enzymes in two separate dissolve stages is permitted, then the identification complexity of $X$ with respect to $\mathcal{C}$ is $O\left(\frac{\left|\pi_X\right|}{\log \left| \pi_X \right|}\right)$.
\end{theorem}

The proof idea of Theorem~\ref{non_planar_two_dissolve_theorem} is similar to that of Theorem~\ref{non_planar_more_shapes} in that we assemble a frame that accepts an input shape $Y$ and allows a border of DNA tiles to assemble if and only if the $Y = X$. In order to overcome the assumption that the $Y$ must have a perimeter-rectangle decomposition, we use two dissolve stages.  Please see Section \ref{two_dissolve_details} for more details of this construction.

%% file: conclusion.tex
%\section{Conclusion}
%\label{conclusion}
%In this paper, we introduced the \emph{shape identification problem} in algorithmic self-assembly. We first studied the complexities associated with identifying $n \times n$ squares with respect to the set of all hole-free shapes using planar tile assembly systems. We essentially found that standard tile complexity coincides exactly with identification complexity with respect to the set of all $n \times n$ squares. We also studied the non-planar identification complexity of a broad class of shapes that have a certain kind of rectangle decomposition or within a model in which there are multiple types of RNA tiles and RNAse enzymes available.

\section{Open Questions}
There are a number of open problems related to using self-assembly for shape identification. First and foremost, a drawback to all of our constructions is that they utilize a system temperature of $\tau = 4$. Although solving the shape identification problem at temperature $\tau = 2$ is impossible by the way the problem is currently formulated, it would be nice to know if there is a solution with temperature $\tau = 3$. Regardless of the system temperature, is it possible to efficiently identify arbitrary hole-free shapes with respect to the set of all hole-free shapes (this is perhaps one of the strongest possible shape-identification results one could hope for)? %Also, is it possible to prove a ``planar'' version of Theorem~\ref{non_planar_technical_theorem}? Speaking of planarity: exactly what class of shapes can be efficiently identified using planar tile assembly systems and what is the complexity of doing so? Note that our planar construction for identifying squares can be easily adapted to identify rectangles. 
Another interesting research direction might be to study the complexities of identifying various classes of shapes in other models of tile-based self-assembly that allow for the \emph{removal} of groups of tiles (e.g., kinetic tile assembly model \cite{Winfree98simulationsof}, multiple temperature model \cite{KS07,AGKS05}, negative glue-strength model \cite{Winf98,ReifSahuYin2006selfdestruct,DotKarMas10}, time-dependent glue-strength model \cite{Sahu05aself}, etc.). 

%% file: appendix.tex
\section{Appendix}

\subsection{Informal Description of the Two-Handed Abstract Tile Assembly Model}
\label{two-handed}
In this subsection we informally review a variant of Erik Winfree's abstract Tile Assembly Model \cite{Winf98,Winfree98simulationsof} modified to model unseeded growth, known as the \emph{two-handed} aTAM, which has been studied previously under various names \cite{AGKS05,DDFIRSS07,Winfree06,Luhrs08,AdlCheGoeHuaWas01,Adl00,Replication10}. In the two-handed aTAM, any two assemblies can attach to each other, rather than enforcing that tiles can only accrete one at a time to an existing seed assembly.

\begin{figure}[htp]
\centering
    \includegraphics[width=.5in]{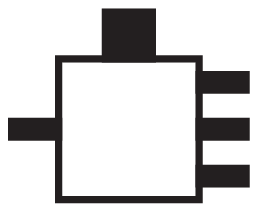} \caption{\label{fig:exampletile} The north glue of this tile has strength 4, the west glue has strength 1, the east glue has strength 3 and the south glue has strength 0.}
\end{figure}

A \emph{tile type} is a unit square with four sides, each having a \emph{glue} consisting of a \emph{label} (a finite string) and \emph{strength} (a natural number).  We represent tiles as squares. Notches on the sides of tile types represent the glue strength of that side. The thick notches represent strength $4$, and otherwise each single notch contributes a strength of one to the glue strength of that side. See Figure~\ref{fig:exampletile} for an example of our tile notation.

We assume a finite set $T$ of tile types, but an infinite number of copies of each tile type, each copy referred to as a \emph{tile}. A \emph{supertile} (a.k.a., \emph{assembly}) is a positioning of tiles on the integer lattice $\mathbb{Z}^2$. Two adjacent tiles in a supertile \emph{interact} if the glues on their abutting sides are equal and have positive strength. Each supertile induces a \emph{binding graph}, a grid graph whose vertices are tiles, with an edge between two tiles if they interact. The supertile is \emph{$\tau$-stable} if every cut of its binding graph has strength at least $\tau$, where the weight of an edge is the strength of the glue it represents. That is, the supertile is stable if at least energy $\tau$ is required to separate the supertile into two parts. A \emph{tile assembly system} (TAS) is a pair $\mathcal{T} = (T,\sigma,\tau)$, where $T$ is a finite tile set, $\sigma$ is an initial seed configuration and $\tau \in \mathbb{N}$ is the \emph{temperature}. Given a TAS $\mathcal{T}=(T,\sigma,\tau)$, a supertile is \emph{producible} if either it is a single tile from $T$, $\sigma$, or it is the $\tau$-stable result of translating two producible assemblies. A supertile $\alpha$ is \emph{terminal} if for every producible supertile $\beta$, $\alpha$ and $\beta$ cannot be $\tau$-stably attached. A TAS is \emph{directed} (a.k.a. \emph{deterministic} or \emph{confluent}) if it has only one terminal, producible supertile and all producible assemblies are finite. Given a connected shape $X \subseteq \mathbb{Z}^2$,  a TAS $\mathcal{T}$ \emph{produces $X$ uniquely} if every producible, terminal supertile places tiles only on positions in $X$ (appropriately translated if necessary).

\subsection{RNA Tiles and the RNAse Enzyme}
\label{subsection_rna_model}
In this paper, we assume that each tile type is defined as being either composed of DNA or of RNA.  By careful selection of the actual nucleotides used to create the glues, tile types of any combination of compositions can bind together. However, the utility of RNA-based tile types comes from that fact that, at prescribed points during the assembly process, the experimenter can add an RNAse enzyme to the solution which causes all tiles composed of RNA to dissolve.  We assume that when this occurs, all portions of all RNA tiles are completely dissolved, including glue portions that may be bound to DNA tiles, returning the previously bound edges of those DNA tiles to unbound states.

In other words, for a given supertile $\alpha$ that is stable at temperature $\tau$, when the RNAse enzyme is added to the solution, all positions in $\alpha$ which are occupied by RNA tiles become undefined (locations at which no tiles exist). The resultant supertile may not be $\tau$-stable and thus defines a multiset of subsupertiles consisting of the maximal stable supertiles of $\alpha$ at temperature $\tau$, which we denote as $BREAK^{\tau}(\alpha)$.

Unless explicitly stated, in this paper we subscribe to the restriction that the RNAse enzyme \emph{must} be added exactly once--and \emph{only} after an initial phase of two-handed self-assembly (involving both DNA and RNA tiles at temperature $\tau$) reaches some (intermediate) terminal state. Of course, after the RNAse enzyme has completely dissolved all of the RNA tiles, self-assembly of only DNA tiles is allowed to proceed until a final terminal state is reached. We also assume that tile types \emph{cannot} be added at any point of the self-assembly process, whence all of our constructions in this paper have $O(1)$ \emph{stage complexity}.

The reader is encouraged to consult \cite{Replication10} for a thorough discussion of the RNAse enzyme self-assembly model.

\subsection{Proof Sketch of Theorem~\ref{theorem_square_log_n}}
\label{planar_construction_sketch}
\begin{proof}[sketch]
Let $6 < n \in \mathbb{N}$. We will describe our construction in terms of Figure~\ref{fig:log_construction_overview}. It is important to note that every (super)tile attaches in a planar manner.

\textbf{Verification of the North, West and South Sides of Input Shape:} First, the red tiles form a $4$-stable supertile which binds to the upper-left corner of the input shape. Note that two-handed assembly is required for this to happen since, in the shape identification problem, we cannot assume that the corners of the input shape are marked in any special way. Then the orange tiles count from $2^{\lfloor \log n \rfloor + 1}-n$ up to  $2^{\lfloor \log n \rfloor + 1}-1$ using a slightly modified version of an \emph{optimal binary counter} used in \cite{AdChGoHu01}. In this construction, we encode $2^{\lfloor \log n \rfloor + 1}-n$ using $\left\lfloor\log \left(2^{\lfloor \log n \rfloor + 1}-n\right)\right\rfloor + 1 = O(\log n)$ unique tile types that assemble a right-triangle of orange tiles whose topmost row encodes the number $2^{\lfloor \log n \rfloor + 1}-n$.

The last (topmost) row of orange tiles detects the end of the count. The green ``filler'' tiles can then fill in the corner that is created by the two groups of orange tiles (the leftmost west-growing group of orange tiles is simply a rotated version of the counter described above). While the orange and green tiles are assembling, the yellow tiles assemble a rectangle of height $n + \lfloor \log n \rfloor + 2$ and width $n$ on top of the input shape (this might not be possible if the input shape is not an $n \times n$ square). If the topmost row of orange tiles is flush with the topmost row of the yellow tiles, then a row of blue tiles assembles on top of the yellow tiles (from left to right) until the rightmost column of the yellow tiles is reached. At this point, the blue tiles assemble back to the left to meet up with the oppositely-oriented group of (north-growing) blue tiles associated with the left side of the input shape.  If this meeting occurs, then the blue ``YES'' tile attaches and waits for the third single-tile-wide path of blue tiles associated with the south side of the input shape to assemble. If all three sides of the input shape agree, then a ``YES'' indigo tile initiates the assembly of a row of indigo tiles that grow back to the right across the top of the assembly and eventually down the right side of the (north-growing) yellow tiles until encountering the row of ``G'' yellow tiles.

\textbf{Assembly of the Border:} While the self-assembly described in the previous paragraph is taking place, the red ``lower-right'' corner tiles attach to the input shape (of course, this might not happen if the input shape is not a square). Then, and only once the downward-growing indigo tiles (mentioned above) reaches the yellow ``G'' row of tiles, a group of east-growing indigo tiles counts from $2^{\lfloor \log n \rfloor + 1}-3 - 1$ up to  $2^{\lfloor \log n \rfloor + 1}-1$ using the same modified optimal counting scheme as above. We encode $2^{\lfloor \log n \rfloor + 1}-3 - 1$ using $\left\lfloor \log\left(2^{\lfloor \log n \rfloor + 1}-3 - 1\right)\right\rfloor + 1 = O(\log n)$ unique tile types. When the indigo counter reaches its maximum value, a group of violet tiles assembles so that its left edge is $n-3$ tiles tall and whose top edge is flush with that of the input shape. As the violet tiles are assembling, the grey ``U'' tiles attach to the input shape one at a time in a planar fashion. The main job of the ``U'' tiles is to determine if the right side of the input shape is consistent with that of an $n \times n$ square--if it is, then the input shape must be a square. See Figure~\ref{fig:construction_sequence} for an illustration of the order of assembly in our construction.  Note that, independent of the value of $n$, only the last two (the southernmost) tiles to be placed in this column will be of the types ``U2'' and ``U3'', and there may be many more ``U'' tiles.  These final two tiles are necessary to stabilize the entire column after RNAse is added and can only be placed if the east side, which is the final side to be checked, also matches that of an $n \times n$ square.

\begin{figure}[htp]
\centering
    \subfloat[][All of these groups of tiles can assemble in parallel]{%
        \label{fig:construction_sequence_5}%
        \includegraphics[width=1.5in]{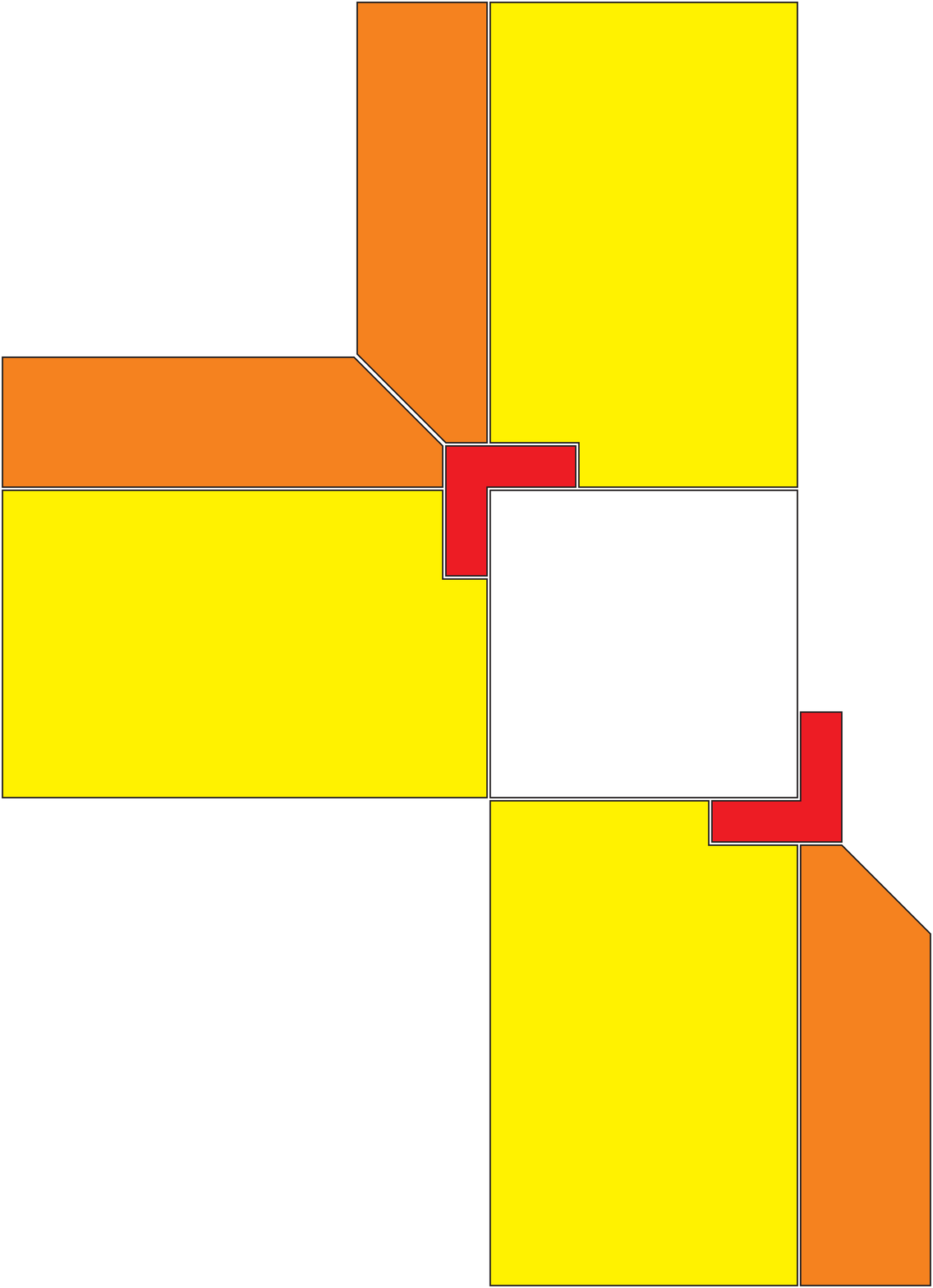}}
        \hspace{20pt}%
    \subfloat[][The left and top borders ``agree'' on having the same dimension via the blue paths of tiles]{%
        \label{fig:construction_sequence_6}%
        \includegraphics[width=1.64in]{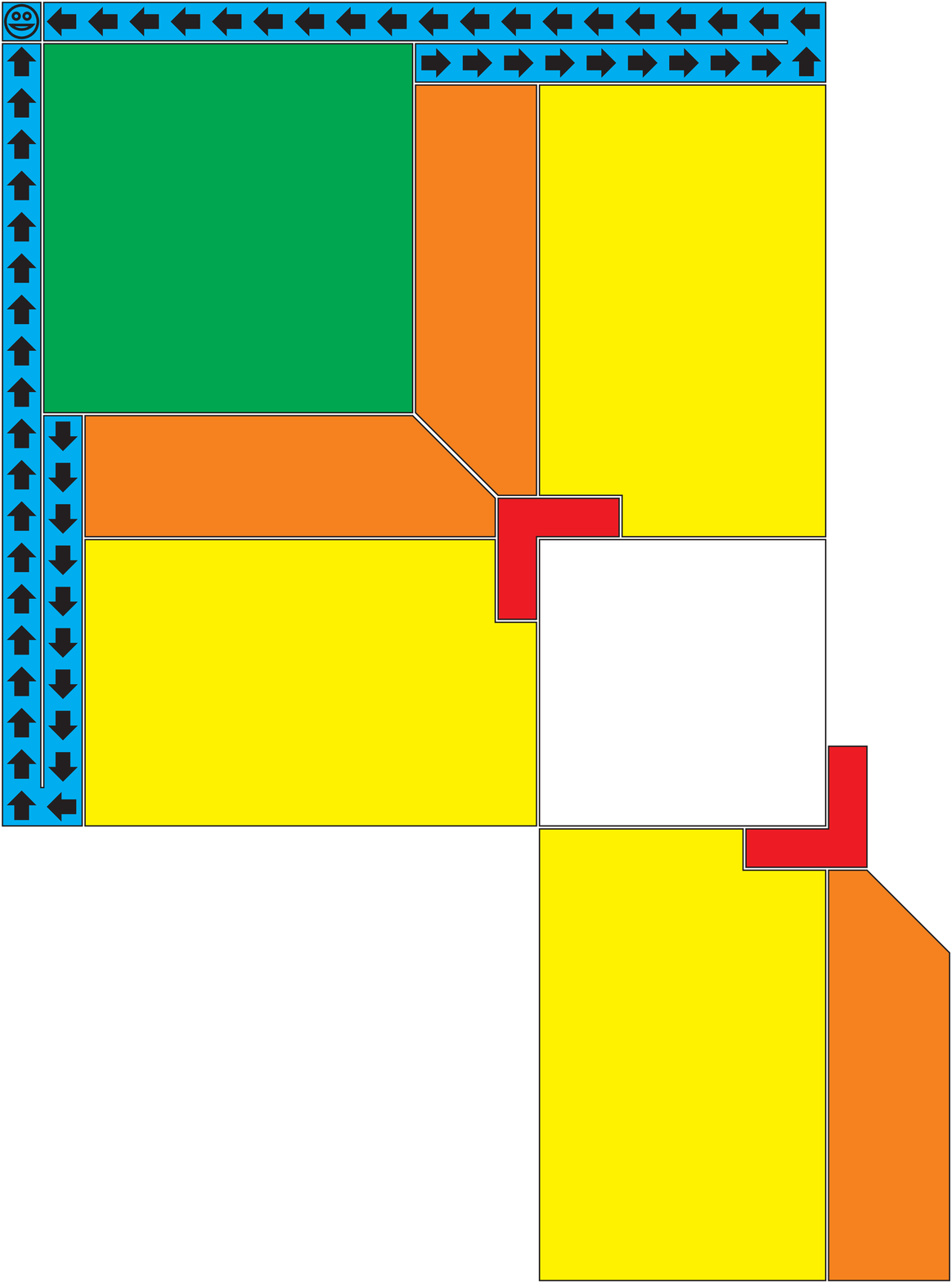}}\\
    \subfloat[][The bottom border ``agrees'' with whatever dimension the top and left borders agreed upon via a final blue path of tiles]{%
        \label{fig:construction_sequence_7}%
        \includegraphics[width=1.72in]{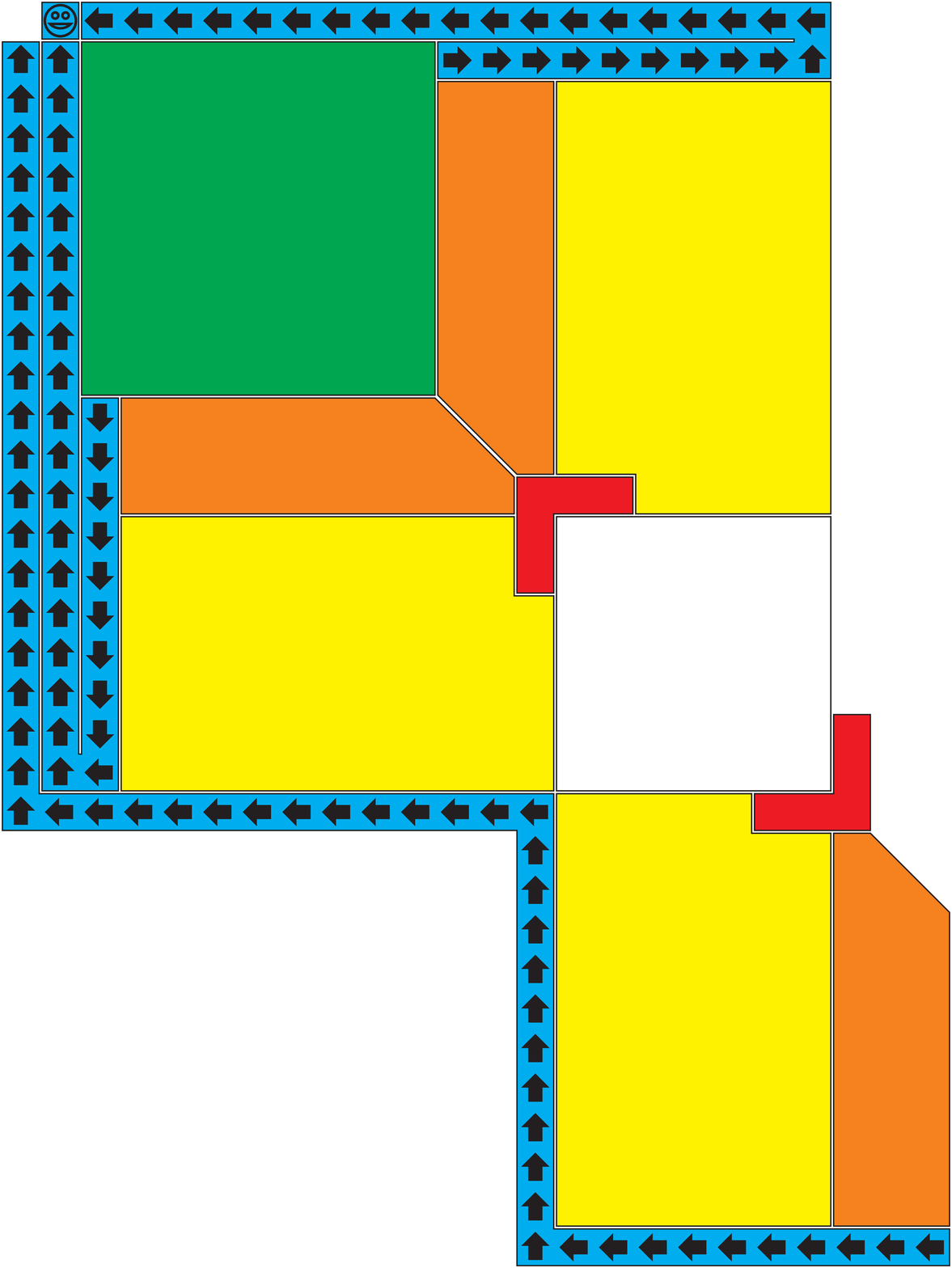}}
        \hspace{20pt}%
     \subfloat[][After (and only if) the left, top and bottom borders agree on having the same dimension, the indigo path initiates the violet group of tiles that determine if the right side of the input shape is consistent with a square.]{%
        \label{fig:construction_sequence_8}%
        \includegraphics[width=2.14in]{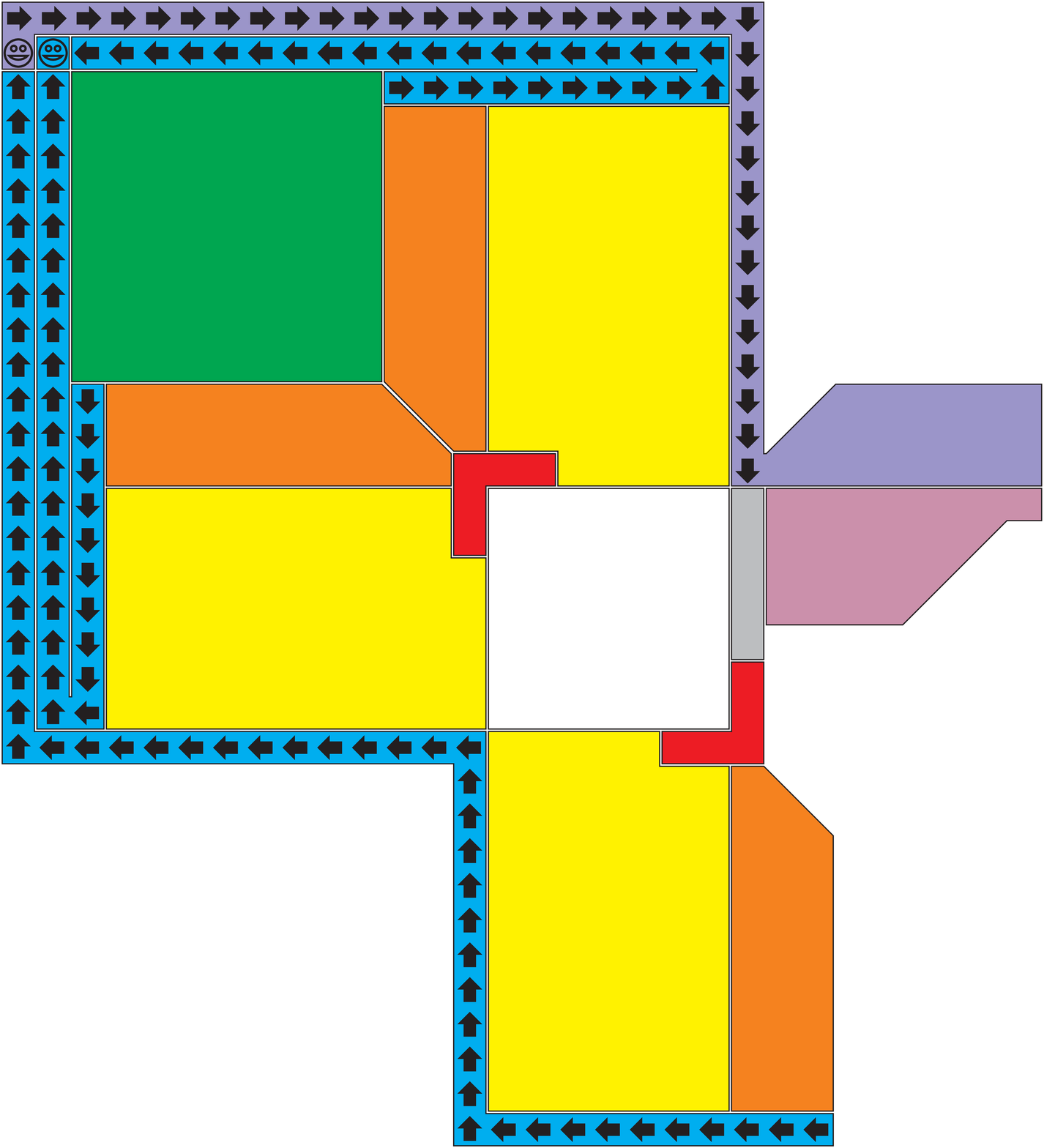}}%
    \caption{\label{fig:construction_sequence} The order of assembly is Red, Orange, Yellow, Green, Blue, Indigo and Violet. The little black arrows represent the order of assembly of single-tile-wide paths.}%
\end{figure}

After the ``U2'' and ``U3'' tiles attach to the right border of the input shape, this stage of the assembly is terminal and we add the RNAse enzyme to dissolve all of the RNA tiles (those that are not part of the input shape or have labels containing ``U''). Finally, the remaining grey tiles attach to the left, top and bottom borders of the input shape as shown in Figure~\ref{fig:log_construction_overview_final}\subref{fig:log_construction_overview_final_2} in a clockwise fashion.

\begin{figure}[htp]
\centering
    \subfloat[][Right after the RNAse enzyme is added.]{%
        \label{fig:log_construction_overview_final_1}%
        \includegraphics[width=2.12in]{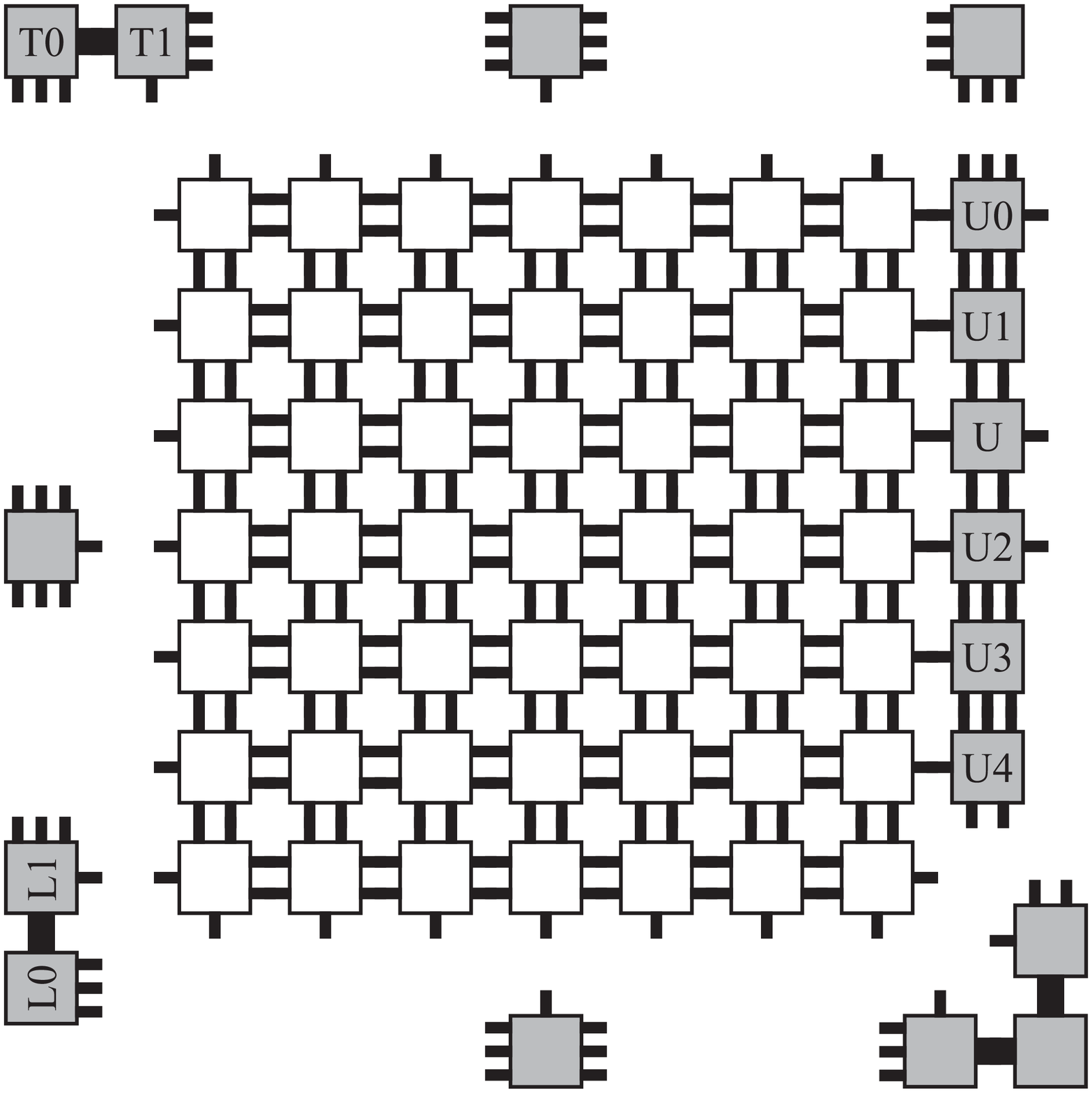}}%
        \hspace{20pt}%
     \subfloat[][Eventually, the grey DNA tiles assemble around the border.]{%
        \label{fig:log_construction_overview_final_2}%
        \includegraphics[width=2.0in]{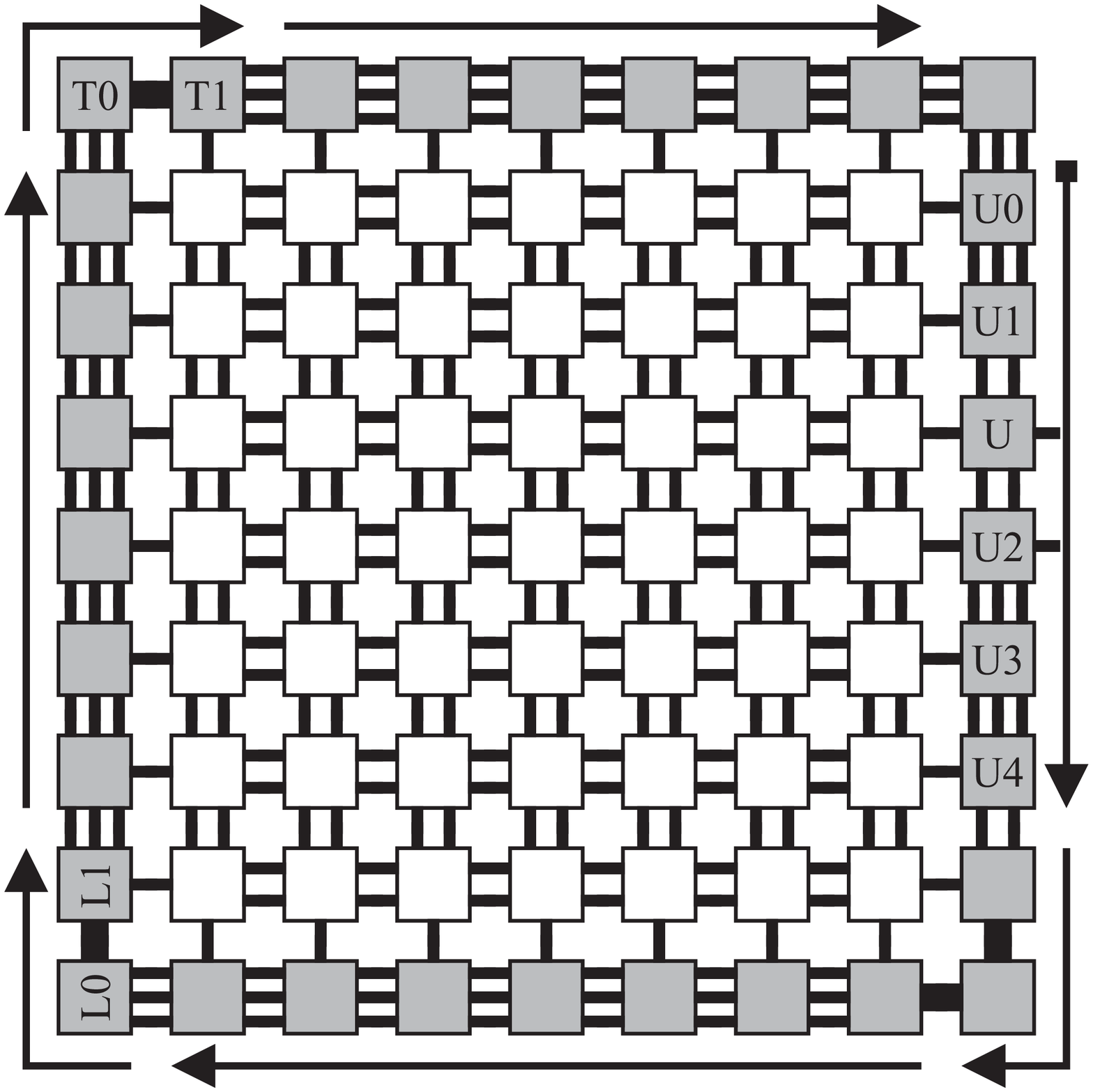}}%\\
    \caption{\label{fig:log_construction_overview_final} This is what happens after we add the RNAse enzyme to the assembly from Figure~\ref{fig:log_construction_overview}.}%
\end{figure}

Note that if the input shape does not match the target square, then there are several situations that might occur. First, the topmost orange tile will either be too high (or not high enough), whence it will not be able to cooperate with the leftmost tile in the top row of the north-growing yellow tiles. This means that the rows of blue tiles will never cooperate and ``agree'' that three of the sides of the input shape are actually consistent with those of an $n \times n$ square. As a result, when the RNAse enzyme is added, there will be no ``U'' tiles attached to the border of the input shape. Second, if the right border of the input shape is not consistent with the target square, then the grey ``U'' tiles will search for the point at which the right border becomes inconsistent with a square. This will hinder any further ``U'' tiles to attach past this point. When the RNAse enzyme is added, the previously attached ``U'' tiles will not bind with sufficient strength and will ultimately fall off of the input shape. This situation is illustrated in Figure~\ref{fig:wrong}.
\begin{figure}[htp]
\centering
    \subfloat[][The ``U'' tiles try to attach along the right border.]{%
        \label{fig:wrong1}%
        \includegraphics[width=1.93in]{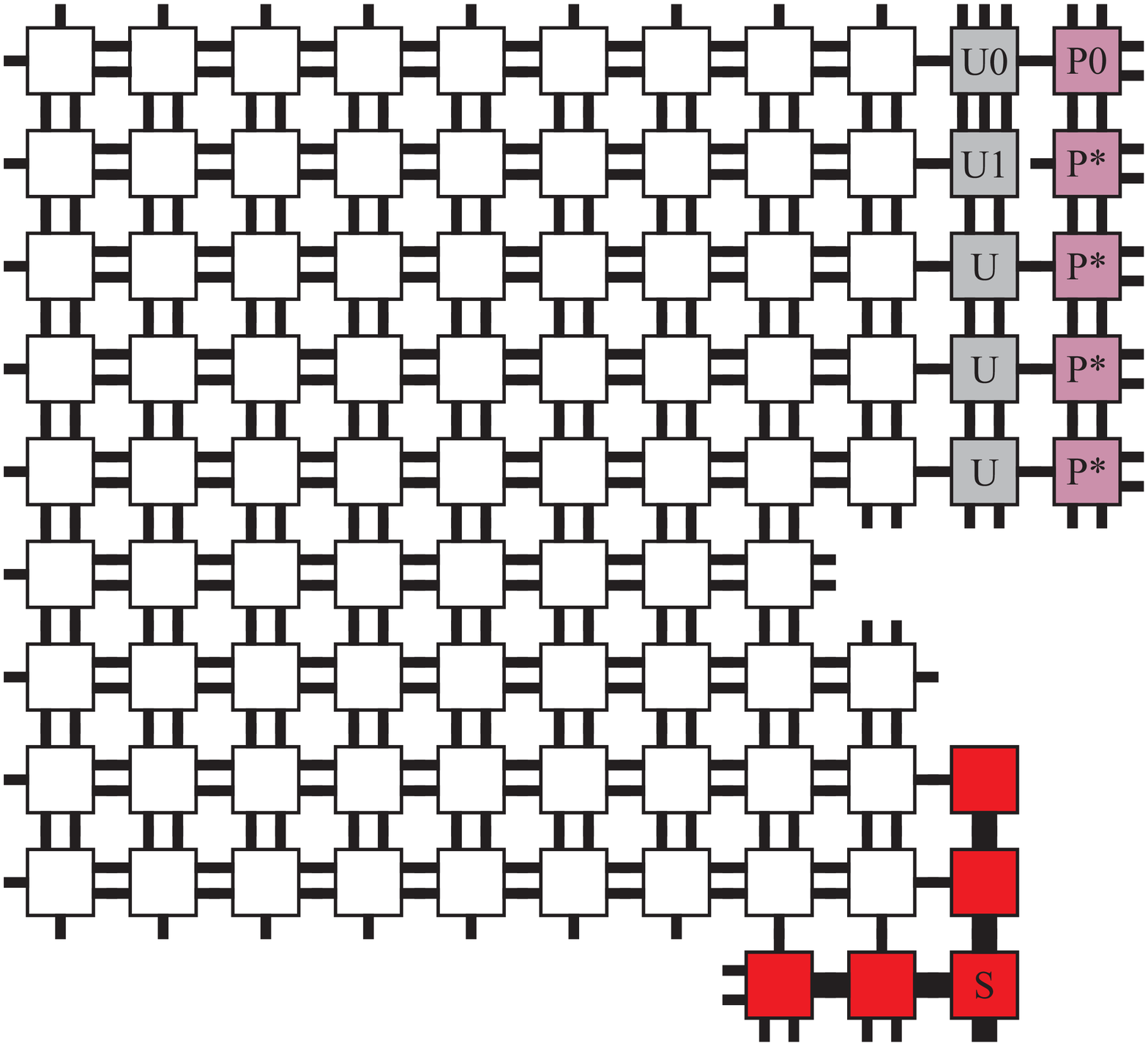}}%
        \hspace{20pt}%
    \subfloat[][After adding the RNAse enzyme: notice that the outlined (bottommost) ``U'' tile binds with strength $3 < \tau = 4$.]{%
        \label{fig:wrong2}%
        \includegraphics[width=1.75in]{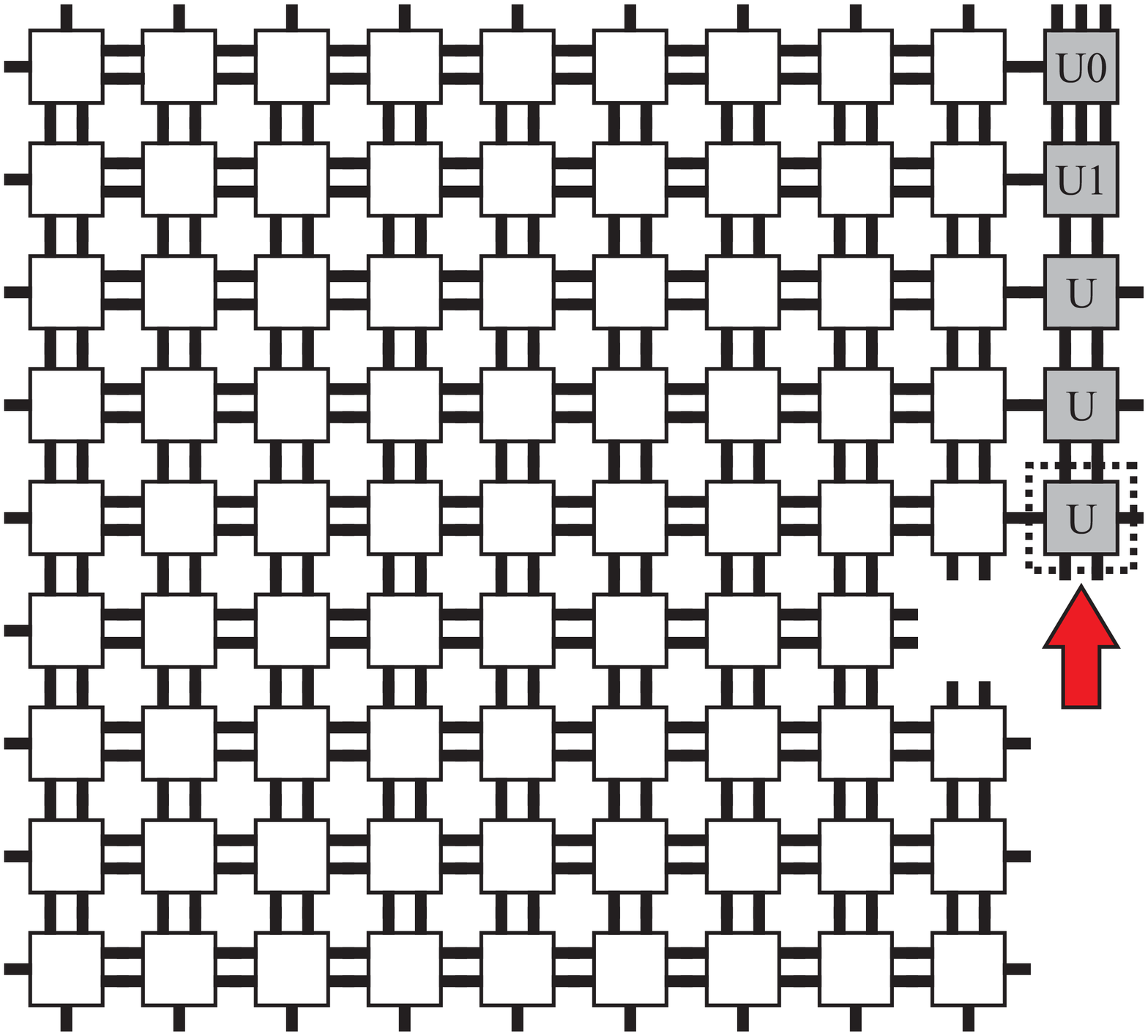}}\\
       % \vspace{20pt}%
     \subfloat[][The ``U'' tiles fall off one at a time until these two outlined ``U'' tiles remain, which collectively bind with total strength $2 < \tau = 4$.]{%
        \label{fig:wrong3}%
        \includegraphics[width=1.75in]{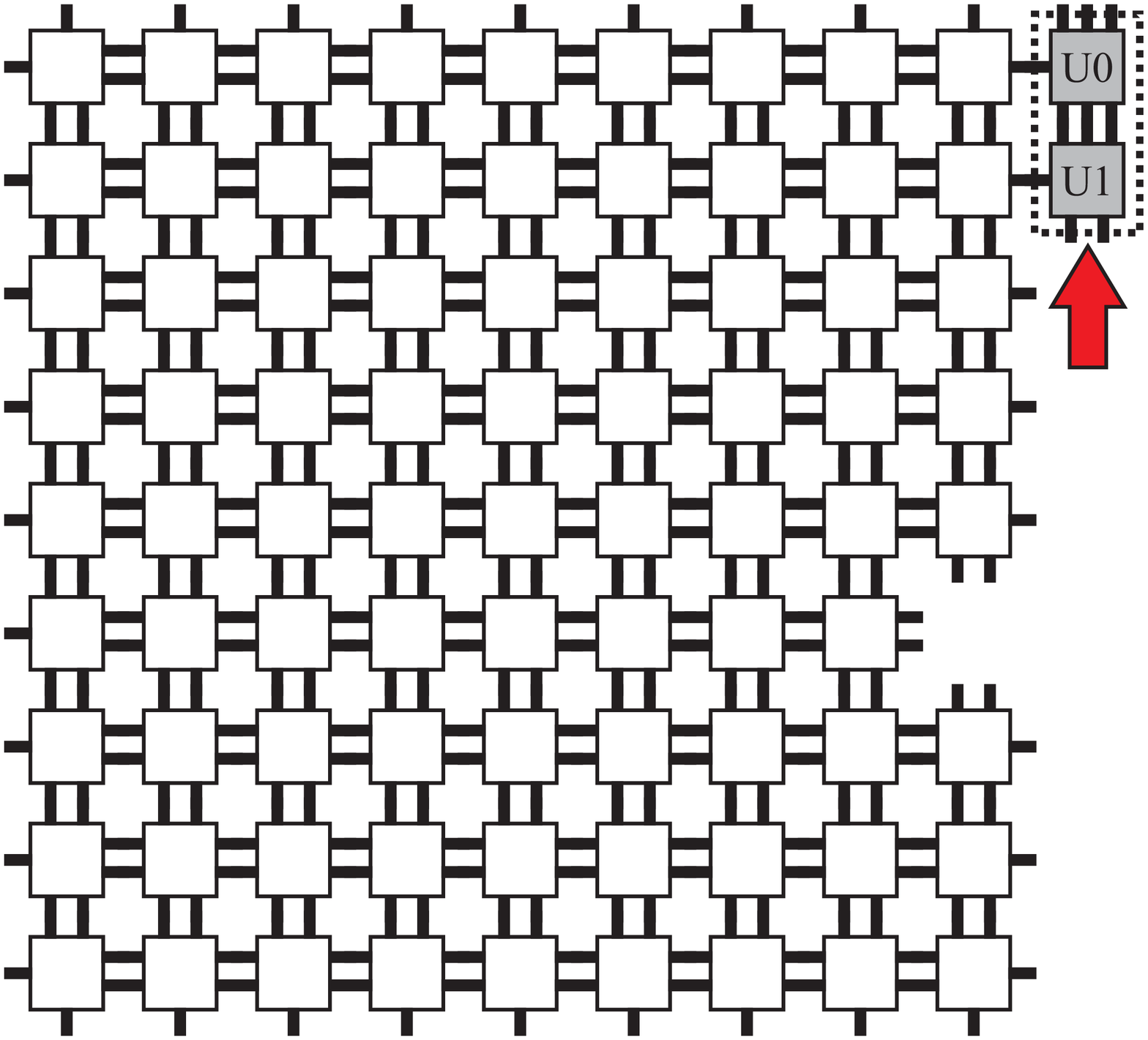}}
    \caption{\small A ``no'' instance of the shape identification problem. Notice that there is a single tile missing in the rightmost column of the input shape making it a non-square input shape.}%
        \label{fig:wrong}%
\end{figure}
\end{proof}

It is worthy of note that our construction for Theorem~\ref{theorem_square_log_n} satisfies a stronger--perhaps more ``experimentally realistic''--formulation of the shape identification problem in which the seed assembly is not a single input shape but instead is a (possibly infinite) \emph{collection} of input shapes and the goal is to correctly identify \emph{all copies} of the target shape.

Finally, for the portions of the construction whose tile complexity was already mentioned, they all required $O(\log n)$ tiles.  For every other component there is a constant sized tile set, independent of $n$.  Therefore, the resulting construction require $O(\log n)$ tile types.

\subsection{Proof Sketch of Theorem~\ref{theorem_square_optimal}}
\label{square_optimal_proof_sketch}

\begin{proof}[sketch]
We will employ an ingenious \emph{optimal encoding scheme} that was introduced by Adleman, Cheng, Goel, and Huang \cite{AdChGoHu01} and then later modified by Soloveichik and Winfree \cite{SolWin07} to work at temperature $\tau = 2$. This method uses $O\left(\frac{n}{\log n}\right)$ unique tile types to encode the bit string $x = x_0x_1\cdots x_{n-1}$, whence Theorem~\ref{theorem_square_optimal} follows. Intuitively, the optimal encoding scheme expresses $x$ as the concatenation of $\lceil n/k \rceil$ strings each of length $k \in \mathbb{N}$, where $k$ is the smallest number satisfying $\frac{n}{\log n} < 2^k$. Each length $k$ substring is encoded as a unique seed tile type that collectively assemble into a seed row of tiles. An ``unpacking'' procedure is carried out in rows of tiles above this initial seed row of (as illustrated in Figure~\ref{fig:unpacking_example}) until the bits of $x$ are fully unpacked. The reader is encouraged to consult \cite{AdChGoHu01,SolWin07} for a detailed analysis of this optimal encoding scheme.

\begin{figure}[htp]
\centering
    \includegraphics[width=3.0in]{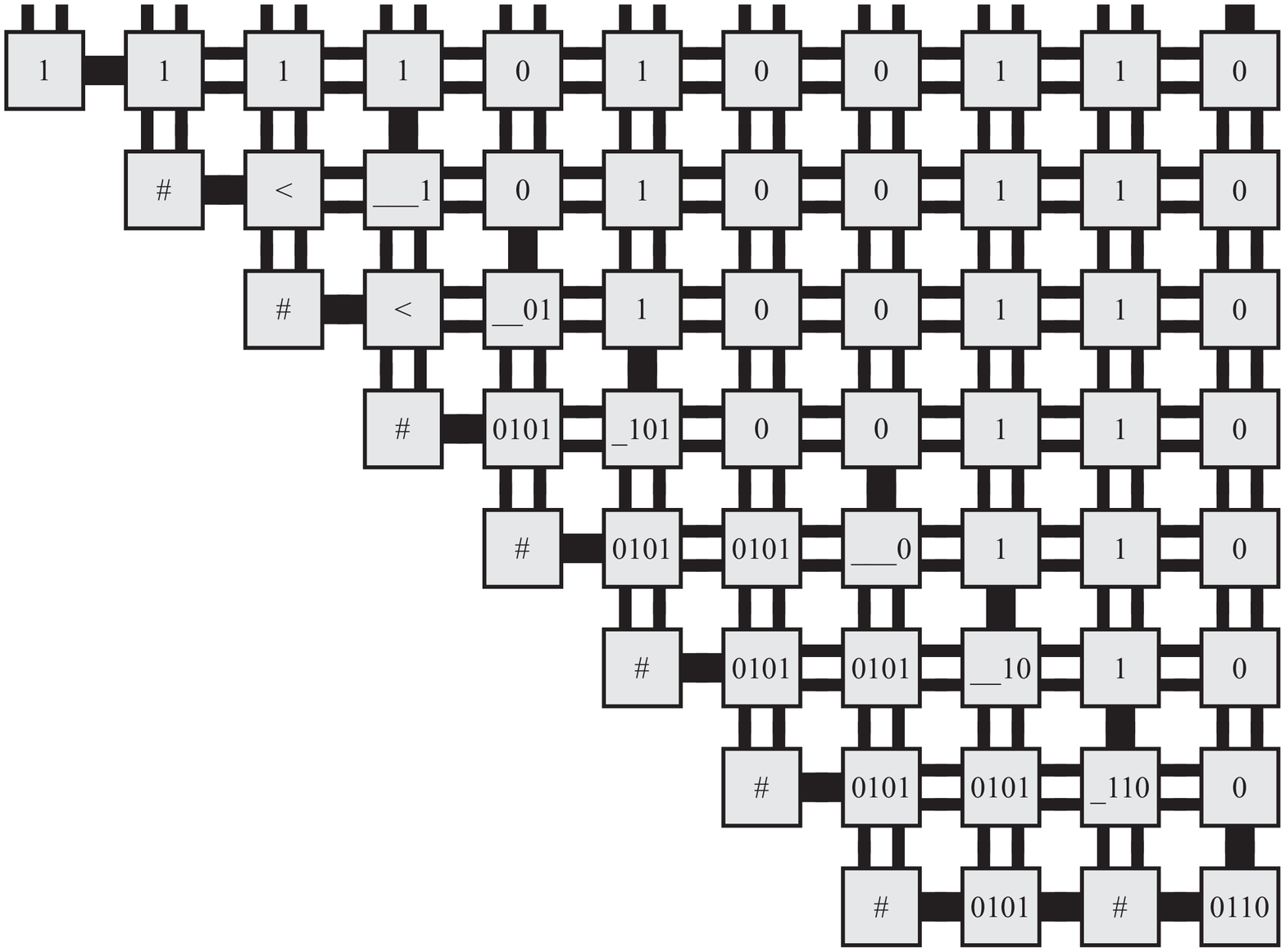} \caption{\label{fig:unpacking_example}  An example of how the string $x = 10100110$ is unpacked. In this example, $k = 4$. Note that this image is essentially Figure 5.7 from \cite{SolWin07} with many of the details omitted. }
\end{figure}
Note that the north glues of the final (topmost) row of the unpacking process each advertise a corresponding bit of the input string $x$ (padded with leading $1$'s). These bits are used to seed all of the modified optimal binary counters in our construction. After this unpacking process completes, the construction for Theorem~\ref{theorem_square_log_n} can proceed normally.
\end{proof}

\subsection{Proof Sketch of Theorem~\ref{universal_theorem}}
\label{universal_squares_proof_sketch}
\begin{proof}[sketch]
We implicitly define our universal tile set $T$ in terms of Figure~\ref{fig:universal_construction_overview}. Now let $6 < n \in \mathbb{N}$. Our construction proceeds as follows.

\begin{figure}[htp]
\centering
    \includegraphics[width=\textwidth]{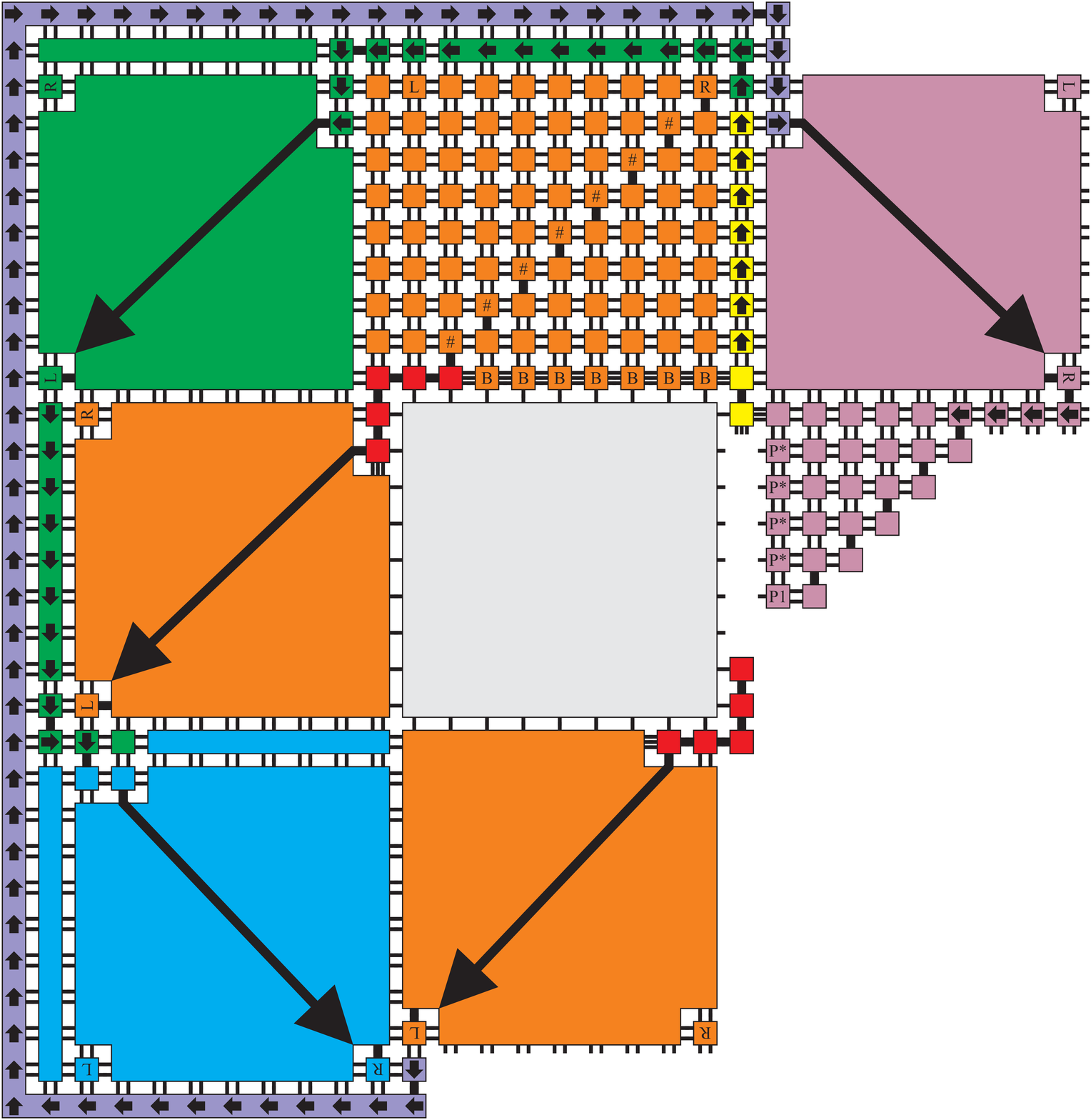} \caption{\label{fig:universal_construction_overview} Overview of universal square identification scheme. Similar to Figure~\ref{fig:log_construction_overview}, the order of assembly is: red, orange, yellow, green, blue, indigo and violet. The border of DNA tiles assembles exactly the same as depicted in Figure~\ref{fig:log_construction_overview_final} except the ``U0'' tile is not used in this construction. }
\end{figure}

The orange unary counters (a.k.a., \emph{shifters}) extract (in unary) the dimensions of the north, west and south sides of the input shape in parallel. The single-tile-wide yellow path of tiles attaches to the upper right corner of the input shape as well as to the right column of the north-growing unary counter. Yellow tiles grow to the north searching for the top of the aforementioned north-growing orange unary counter. When the yellow tiles find the top row, they initiate the self-assembly of the green tiles whose purpose is to search for the column directly to the right of the left most column of the north-growing orange unary counter. Then, the green tiles initiate the growth of a west-growing unary green counter that assembles a square (with dimensions equal to the length of the north side of the input shape) to the left of the north-growing orange unary counter. If the input shape is a square, then the left most tile in the bottom most row of this green (west-growing) unary counter will be positioned precisely one tile above and to the left of the left most tile in the top most row of the west-growing orange unary counter. This allows for a south-growing path of green tiles to proceed until the bottom most row of the west-growing orange unary counter is encountered. Then, a south-growing blue unary counter assembles a square (with dimensions equal to the length of the west side of the input shape) to the south of the west-growing orange unary counter. As before with the west-growing green unary counter, we force the bottom most tile in the right most row of this south-growing blue unary counter to be exactly one tile below and to the left of the left most tile in the bottom most row of the south-growing orange unary counter. In this corner, the growth of a single-tile-wide path of indigo tiles is initiated. This path of indigo tiles assembles along the outside of the entire assembly in a clockwise fashion until reaching the upper right green tiles. When the indigo tiles reach the aforementioned green tiles, an east-growing violet unary counter assembles to the right of the north-growing orange unary counter. When the final (lower right most) violet unary counter attaches, a triangle whose left side dimension is equal to the length of the north side of the input shape--minus three--assembles to the south of the previously assembled east-growing violet unary counter (note that the left most column of this final violet square is flush with the left most column of the violet unary counter to which it attaches thus leaving a single-tile-wide path for DNA border tiles to attach). At this point, DNA border tiles attach exactly as they do in the construction for Theorem~\ref{theorem_square_log_n} with the exception that the ``U0'' tile is never used.
\end{proof}

\subsection{Proof Sketch of Theorem~\ref{non_planar_more_shapes}}
\label{non-planar_proof_sketch}

\begin{proof}[sketch]
Assume that $X$ has a nontrivial perimeter-rectangle decomposition with $n=4$ rectangles--two on the north side and two on the south side of the shape--as the special cases for $n < 4$ are easy to handle.

We define $bin:\mathbb{N} \rightarrow \{0,1\}^*$ as the standard binary (a.k.a., base-$2$) representation of a natural number. Let $R_0,R_1,\ldots,R_{l-1}$ be rectangles whose south edges touch the south edge of $\widetilde{R}(X)$. We can assume without loss of generality that for all $i,j\in\{0,1,\ldots,l-1\}$, $h(R_i) \ne h(R_j)$ whenever $|i-j| = 1$. We will define strings $x_0,x_1,\ldots,x_{l-1}$ over $\Sigma^* = \{0,1\}^*$ that encode the heights of $R_0,R_1,\ldots,R_{l-1}$ (padded with leading $0$'s) respectively. First, define $x_0 = x'_0x''_0$ where $x''_0 = bin\left( h(R_0)-3\right)$ and
$$
x'_0 = \left\{
\begin{array}{ll}
0^{w(R_0)-1- |bin(h(R_0)-3)|} &\textmd{ if } h(R_0) > h(R_1) \\
0^{w(R_0)+1- |bin(h(R_0)-3)|} &\textmd{ otherwise. } \\
\end{array}
\right.
$$
We assume that, for any alphabet symbol $a$ and $q \leq 0$, the expression $a^{q}$ evaluates to the empty string $\lambda$. For $1 \leq i < j < k < l-1$, define $x_j = x'_jx''_j$ where $x''_j = bin\left(h(R_j)-3\right)$ and
$$
x'_j = \left\{
\begin{array}{ll}
0^{w(R_j)+2- |bin(h(R_j)-3)|} &\textmd{ if } h(R_i) > h(R_j) < h(R_k) \\
0^{w(R_j)-2-|bin(h(R_j)-3)|} &\textmd{ if } h(R_i) < h(R_j) > h(R_k) \\
0^{w(R_j)-|bin(h(R_j)-3)|} &\textmd{ otherwise.} \\
%0^{w(R_j)-|bin(h(R_j)-3)|} &\textmd{ if } h(R_i) < h(R_j) < h(R_k) \\
%0^{w(R_j)-1-|bin(h(R_j)-3)|} &\textmd{ if } h(R_i) = h(R_j) > h(R_k) \\
%0^{w(R_j)+1-|bin(h(R_j)-3)|} &\textmd{ if } h(R_i) > h(R_j) = h(R_k) \\
%0^{w(R_j)+1-|bin(h(R_j)-3)|} &\textmd{ if } h(R_i) = h(R_j) < h(R_k) \\
%0^{w(R_j)-1-|bin(h(R_j)-3)|} &\textmd{ if } h(R_i) < h(R_j) = h(R_k) \\
%0^{w(R_j)-2-|bin(h(R_j)-3)|} &\textmd{ otherwise.}
\end{array}
\right.
$$
Finally, define $x_{l-1} = x'_{l-1}x''_{l-1}$ where $x''_{l-1} = bin\left( h(R_{l-1})-3\right)$ and
$$
x'_{l-1} = \left\{
\begin{array}{ll}
0^{w(R_{l-1})+2-|bin(h(R_{l-1})-3)|} &\textmd{ if } h(R_{l-2}) > h(R_{l-1}) \\
0^{w(R_{l-1})-|bin(h(R_{l-1})-3)|} &\textmd{ otherwise. } \\
\end{array}
\right.
$$
We encode the heights of the rectangles whose north edges touch the north edge of $\widetilde{R}(X)$ as the strings $y_0,y_1,\ldots, y_{m-1}$ over $\Sigma^*$ in a similar fashion. Let $f$ be some computable function satisfying $$f(\langle X \rangle) = y_0y_1\cdots y_{m-1} \# bin\left(h\left(\widetilde{R}(X)\right)\right) \# x_0x_1\cdots x_{l-1}.$$
We can assume that the leftmost and rightmost bits of each $y_i$ for $0 \leq i < m$ and $x_i$ for $0 \leq i < l$ are specially marked so that the least and most significant bits of the binary counters which they will ultimately seed can carry unique signals based on whether or not they will help form a convex or concave corner with their neighboring rectangles (doing this will allow the border of DNA tiles to uniquely assemble if and only if the input shape matches the target shape). Note that in our construction we use a special separator character $\#$ to surround the bit string $bin\left(h\left(\widetilde{R}(X)\right)\right)$ so as to distinguish it as the bit string that describes the height of $\widetilde{R}(X)$.

Our construction is broken up into three logical phases: \emph{unpacking}, \emph{frame assembly} and \emph{border assembly}.

\textbf{Unpacking Phase:} In the \emph{unpacking} phase, we encode a program $\pi_X$ that outputs $\langle X \rangle$ using the optimal encoding scheme of Solevichik and Winfree \cite{SolWin07} (also described at a high-level in Section~\ref{section_squares_optimal}) and then use a fixed universal machine $U$ to simulate $\pi_X$--this requires $O\left(\frac{\left| \pi_X \right|}{\log \left| \pi_X \right|}\right)$ unique tile types. Next, on top of that a new Turing machine simulation uses $\langle X \rangle$ as input and computes $f(\langle X \rangle)$, resulting in the topmost row of the assembly (i.e., the final configuration of the Turing machine that computes $f$) encoding all of the information that is necessary to assemble a frame for the target shape $X$ (the blue, green and orange bars in Figure~\ref{fig:non_planar_construction_sequence}\subref{fig:non_planar_construction_overview_1}).
\begin{figure}[htp]
\centering
    \subfloat[][The program $\pi_X$ (the yellow bar) is decompressed into $\langle X \rangle$ and $f(\langle X \rangle)$ is subsequently computed. The blue, green and orange bars collectively represent the string $f(\langle X \rangle)$.]{%
        \label{fig:non_planar_construction_overview_1}%
        \includegraphics[width=1.35in]{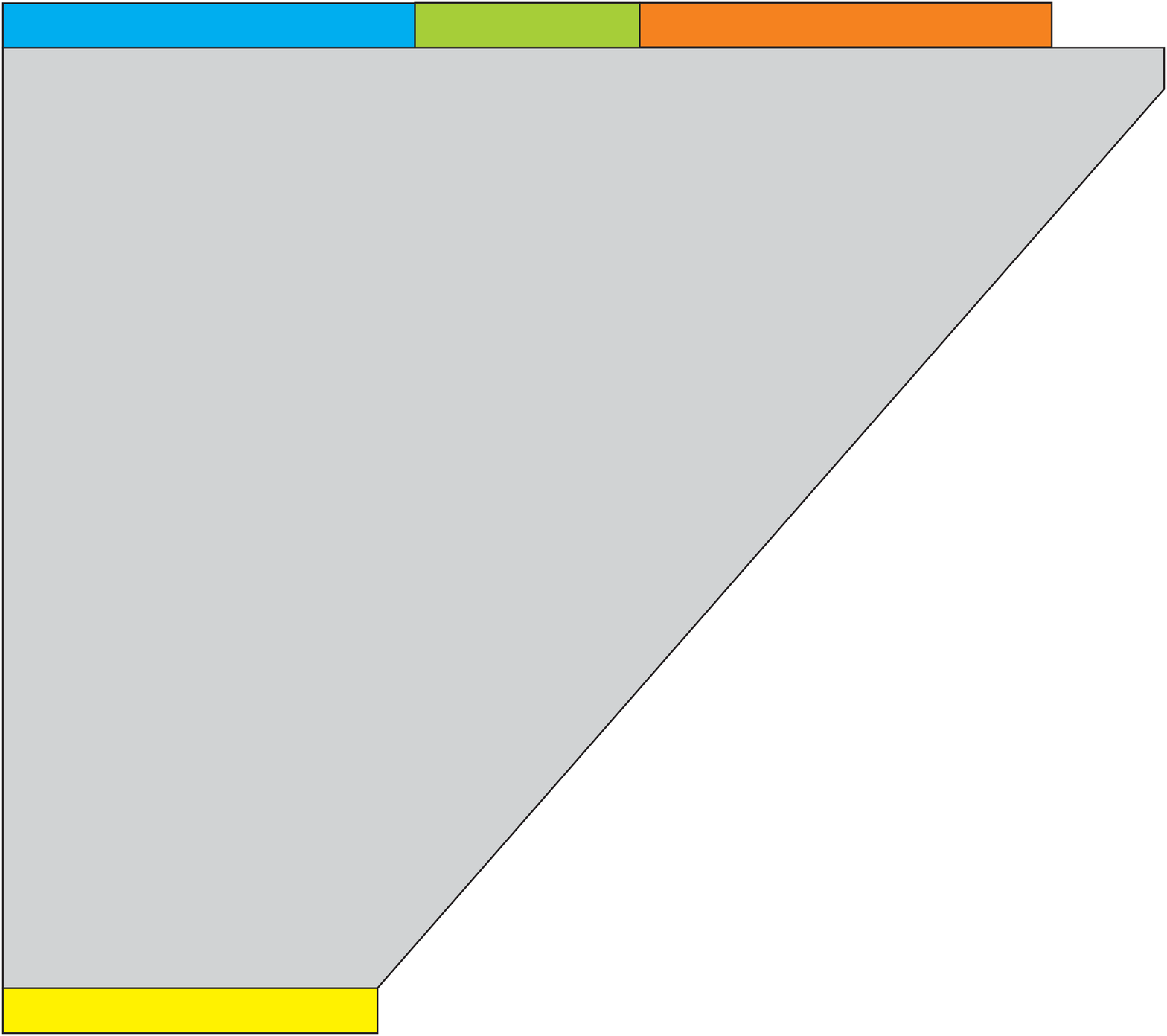}}
        \hspace{20pt}%
    \subfloat[][The portion of the frame that describes the south facing features of the target shape are assembled; the green rectangle is a binary subtractor that assembles $h\left(\widetilde{R}(X)\right)$ rows.]{%
        \label{fig:non_planar_construction_overview_2}%
        \includegraphics[width=1.35in]{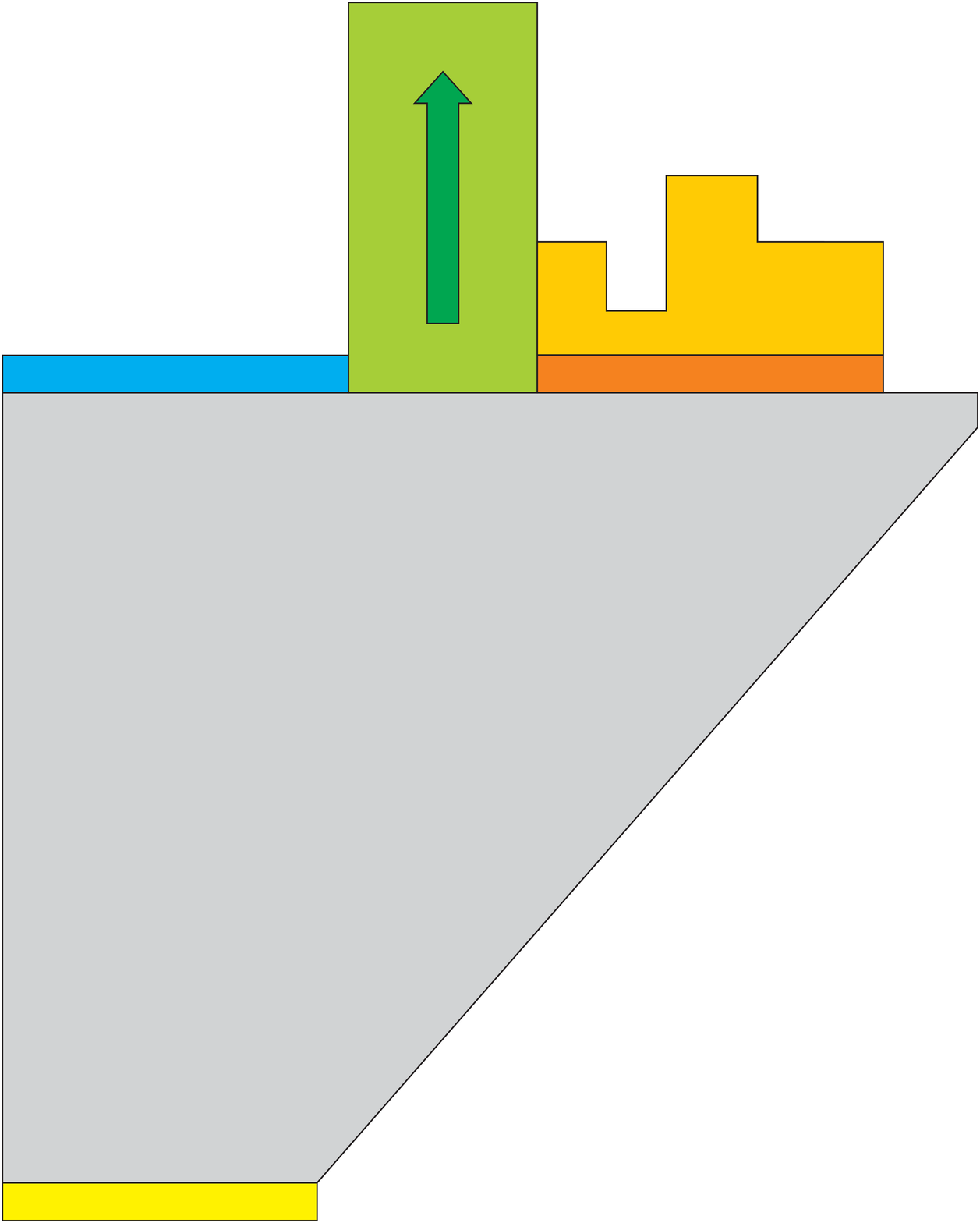}}
        \hspace{20pt}%
     \subfloat[][The information that describes the north-facing features of the target shape are correctly positioned and then assembled.]{%
        \label{fig:non_planar_construction_overview_3}%
        \includegraphics[width=1.35in]{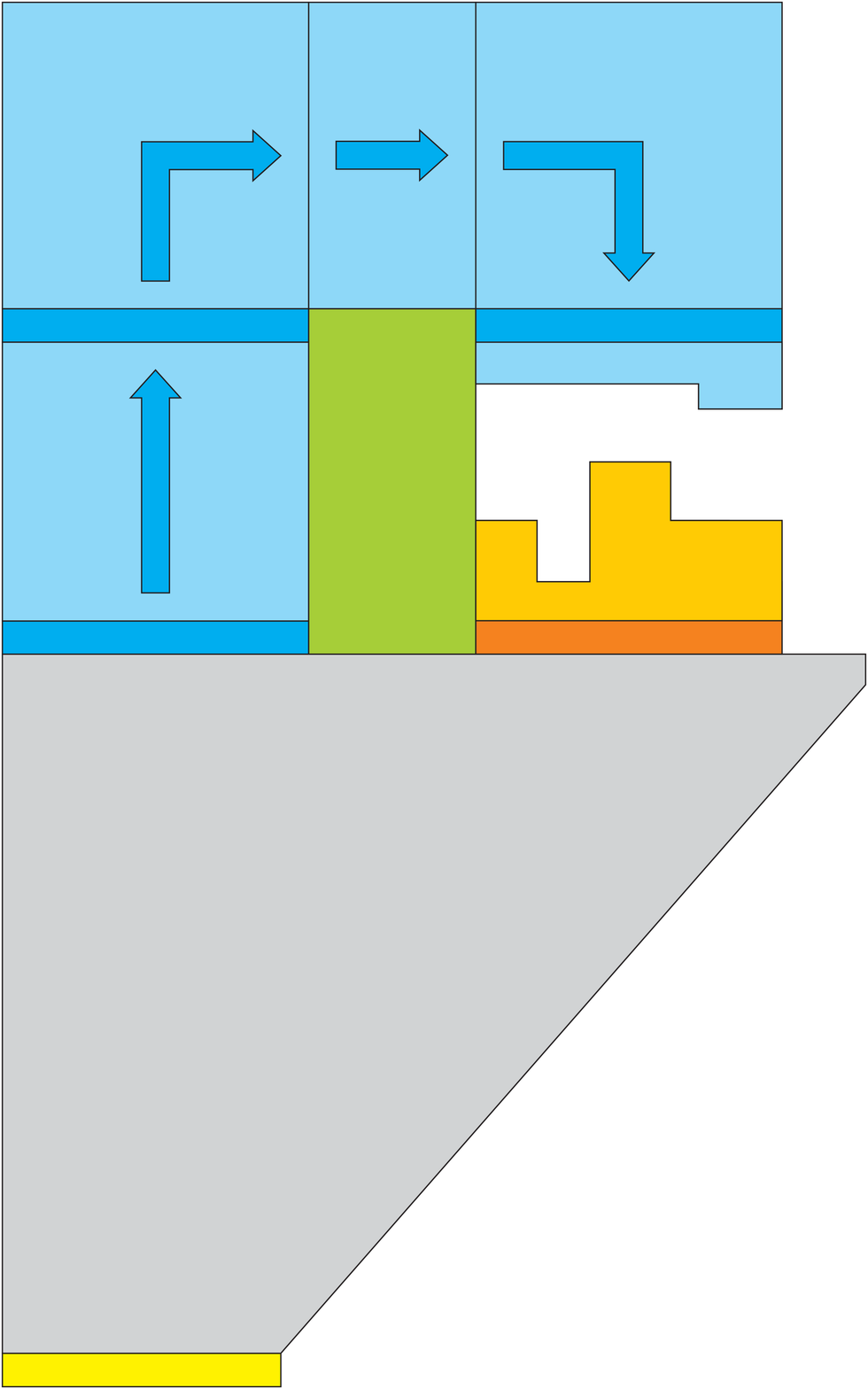}}%
        \hspace{20pt}%
     \subfloat[][Finally, the right border of the frame is assembled and the assembly can correctly identify the target shape.]{%
        \label{fig:non_planar_construction_overview_4}%
        \includegraphics[width=1.35in]{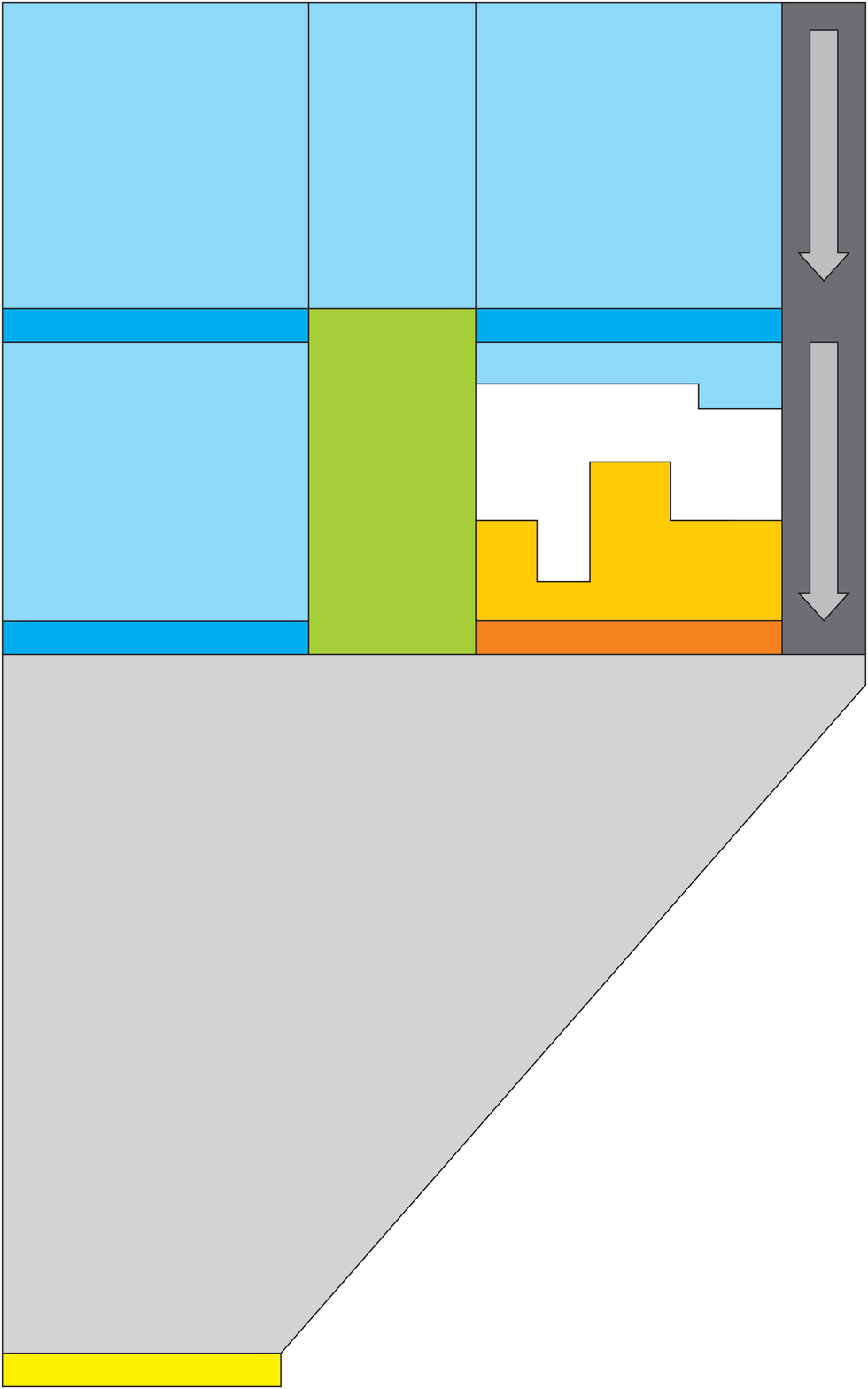}}%
    \caption{\label{fig:non_planar_construction_sequence} Assembly sequence for the unpacking and frame assembly phases.}%
\end{figure}

\textbf{Frame Assembly Phase:} We \emph{assemble the frame} that accepts the target shape $X$ as follows.
\begin{figure}[htp]
\centering
    \includegraphics[width=\textwidth]{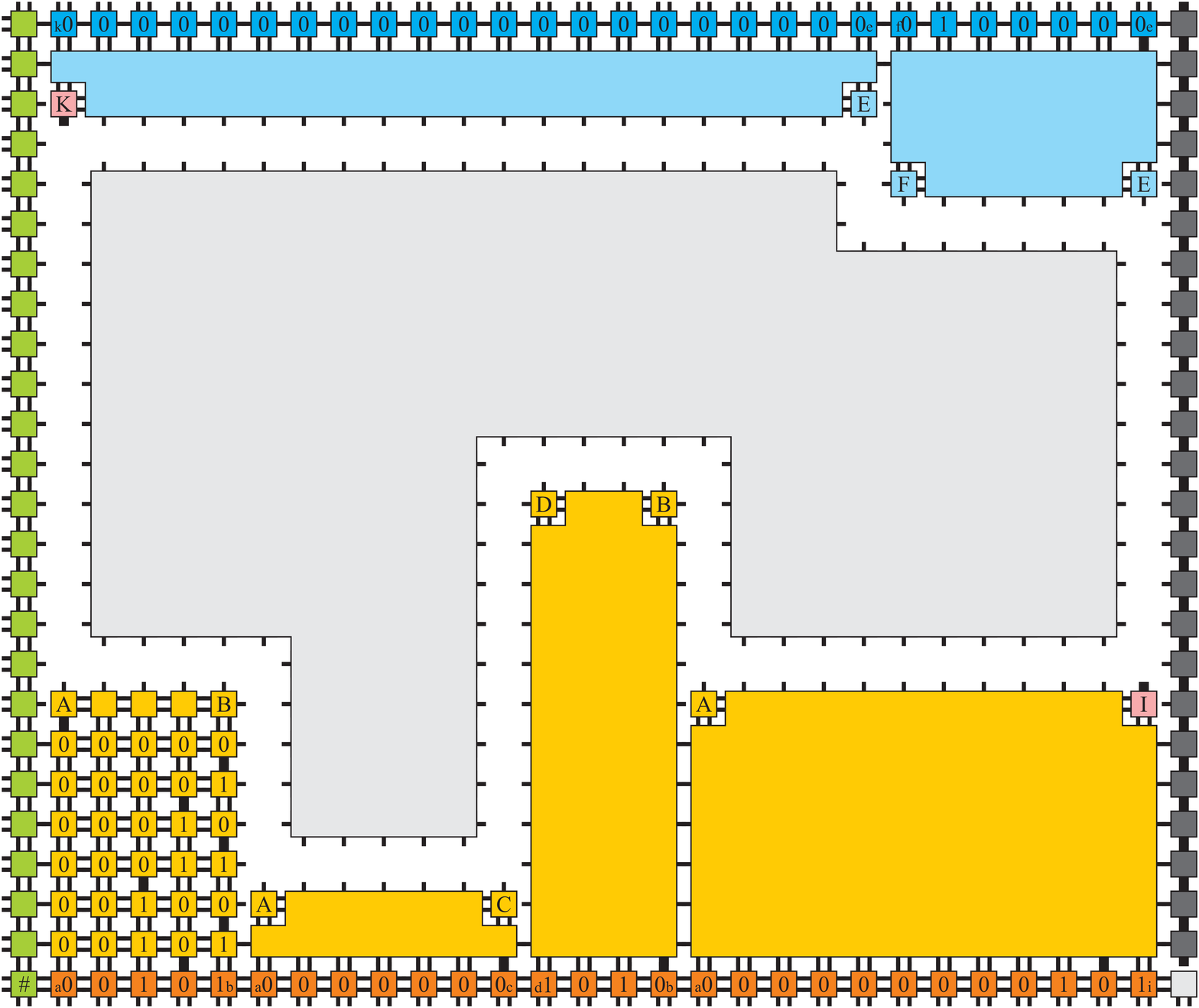} \caption{\label{fig:non_planar_construction_detail} A slightly more detailed version of a portion of Figure~\ref{fig:non_planar_construction_sequence}\subref{fig:non_planar_construction_overview_4}. The target shape $X$ is the grey structure with single strength glues along its border. The light orange rectangles represent binary subtractors (seeded by corresponding portions of dark orange tiles along the bottom row) that assemble portions of the input frame that accept the south-facing features of the target shape. Notice that we mark the left and right corners of the topmost row of each subtractor with tiles that indicate whether a particular corner is concave or convex. We also mark the top left and bottom right most corners with special pink tiles (these tiles will be used in the border assembly phase). South-growing counters that assemble the portion of the frame that accepts the north-facing features of $X$ are constructed similarly.}
\end{figure}
First, the portion of the frame that accepts the south-facing features of $X$ are assembled using a series of north-growing binary subtractors (similar to the optimal binary counter used in \cite{AdChGoHu01}). The final row in each of the subtractors assembles from right to left and attaches appropriate corner tiles that advertise whether or not they form a concave or convex corner with any neighboring subtractors (note that this information was computed by $f$ and is encoded in the strings $x_0,x_1,\ldots x_{l-1}$).

Then, the left border of the frame--simply a wall of tiles whose height is equal to that of $\widetilde{R}(X)$--is constructed via a binary subtractor that assembles $h\left(\widetilde{R}(X)\right)$ many rows. The glues on the east side of every tile in the rightmost column of this subtractor uniformly advertise (to the border tiles which will assemble later) that they are part of the frame.

Next, the portion of the frame that accepts the north-facing features of $X$ is constructed using south-growing binary subtractors similar to how the portion of the frame that accepts the south-facing features was constructed. The information describing these features is first copied up along the left side of the frame and then rotated 180 degrees clockwise so that they are correctly oriented and positioned directly above their north-growing counterparts.

Finally, the right side of the frame is assembled using a single-tile-wide south-growing path of generic frame border tiles that eventually ``bump into'' the portion of the assembly in which the unpacking process was carried out. All of the glues on the west side of these tiles uniformly advertise (to the border tiles which will assemble later) that they are part of the frame. Figure~\ref{fig:non_planar_construction_detail} depicts a fully-assembled frame enclosing a particular target shape.

\textbf{Border Assembly Phase:} In the \emph{border assembly} phase, we must accomplish the task of assembling a complete and fully connected border of DNA tiles if and only if the input shape matches the target shape. Since the system temperature is $\tau = 4$, we can force DNA border tiles to cooperate with (1) the most recently attached border tile (with strength $2$), (2) the frame (with strength $1$) and (3) a single glue on the input shape (with strength $1$). Essentially, if a DNA border tile cannot cooperate with an already-attached border tile, a frame tile and an input shape tile simultaneously, then it will not have sufficient strength to bind--except in the special case when a border tile must attach in a concave corner. The border assembly phase of an instance of the shape identification problem in which the input shape matches the target shape is depicted in Figure~\ref{fig:border_assembly_good}.
\begin{figure}[htp]
\centering
    \subfloat[][The input shape is attached to the frame (with strength $\tau = 4$) via the two purple corner gadgets. The assembly of the DNA border tiles can now proceed in a counter-clockwise fashion starting from the ``L3'' and ``J3'' tiles. ]{%
        \label{fig:border_assembly_good_1}%
        \includegraphics[width=2.50in]{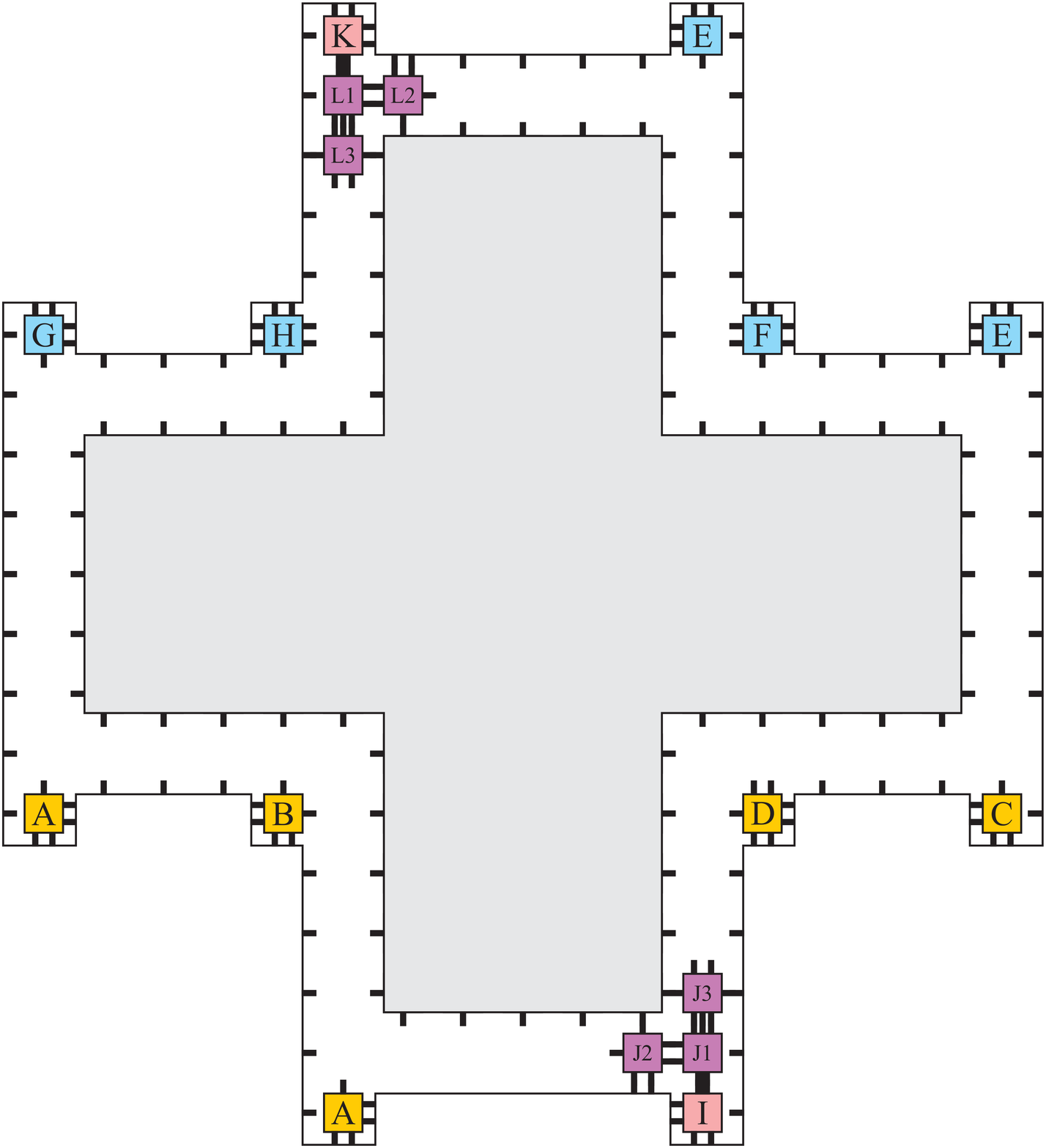}} \\
    \subfloat[][The red DNA border tiles assemble sequentially in a counter-clockwise fashion. Notice that each DNA border cooperates with one bond each from the input shape, a previously attached DNA border tile and the frame (except for corner tiles). This ensures that when the frame is dissolved by the RNAse enzyme, the border will be complete and fully connected.]{%
        \label{fig:border_assembly_good_2}%
        \includegraphics[width=2.50in]{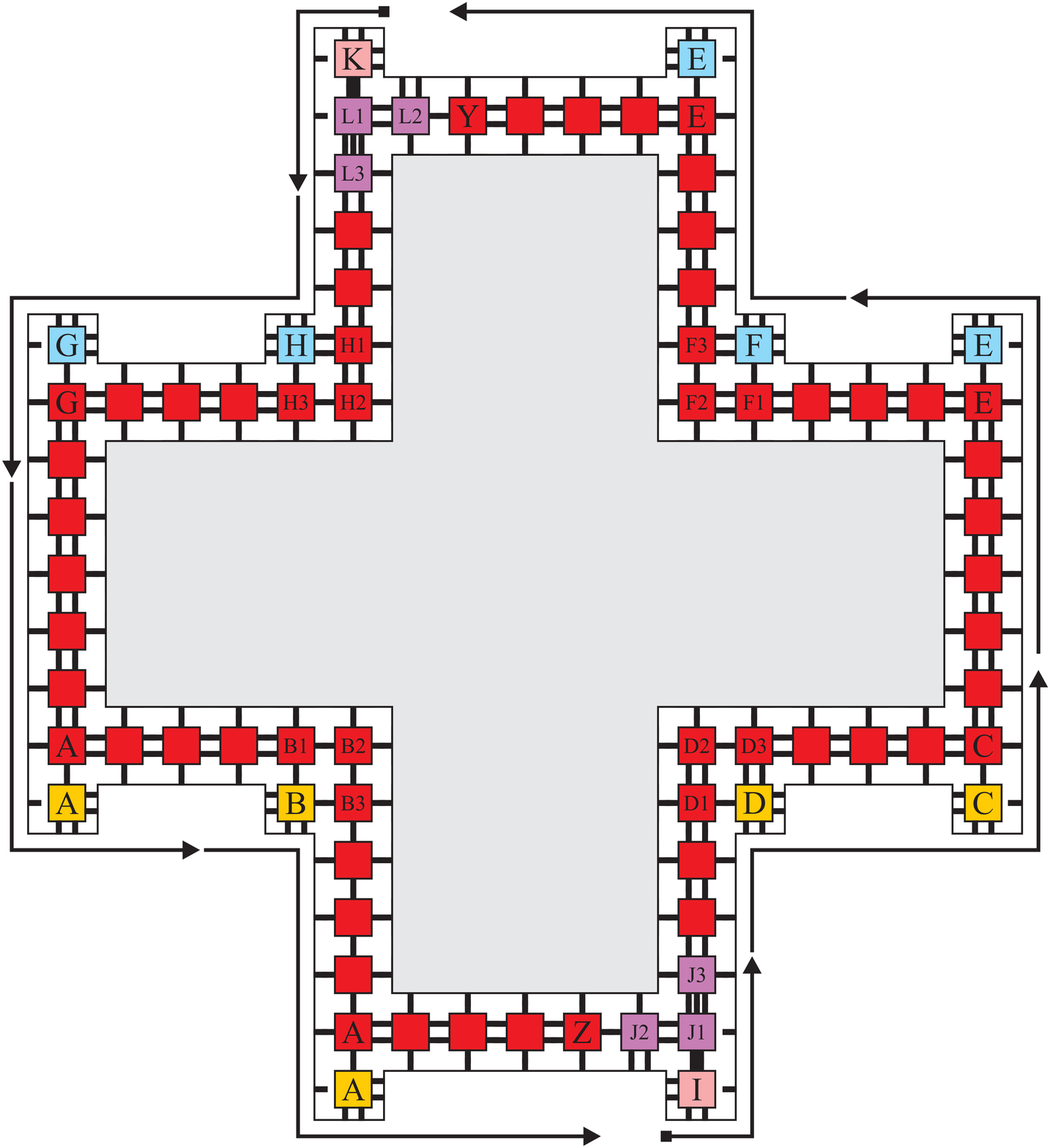}}
    \caption{\label{fig:border_assembly_good} The border assembly phase. The purple and red tiles are DNA tiles. }%
\end{figure}

Following the assembly of the border tiles, the assembly will then be terminal and the RNAse enzyme is added to dissolve away all of the RNA tiles. If the input shape does not match the target shape $X$ then we must ensure that the border of the input shape will eventually be free of any DNA tiles. By the way we enforce that DNA border tiles can attach, we can ensure that, unless the border completely assembles, there is always at least one red or purple DNA border tile that binds to other DNA border tiles and the input shape with strength $3 < \tau = 4$, which means that it will detach from the input shape once the RNA tiles are dissolved. Once this first tile dissociates, then there will exist another DNA border tile that binds with only strength $3$. This process of disassembly of the border one tile at a time (except for purple tiles which can dissociate in groups) will continue until all DNA border tiles are no longer attached to the input shape. This situation is depicted in Figure~\ref{fig:border_assembly_bad}.
\begin{figure}[htp]
\centering
    \subfloat[][In this case, only a portion of the leftmost path of red DNA border tiles can assemble, which means that at least one DNA border tile binds to other border tiles and the input shape with strength $3 < \tau = 4$.]{%
        \label{fig:border_assembly_bad_1}%
        \includegraphics[width=2.51in]{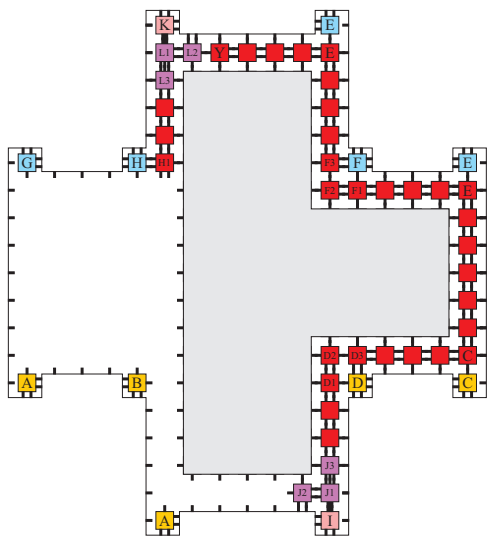}}%
        \hspace{30pt}%
    \subfloat[][After the RNAse enzyme is added and the frame dissolves, the partial border begins to disassemble one tile at a time (except for purple tiles which can dissociate in groups. ]{%
        \label{fig:border_assembly_bad_2}%
        \includegraphics[width=1.75in]{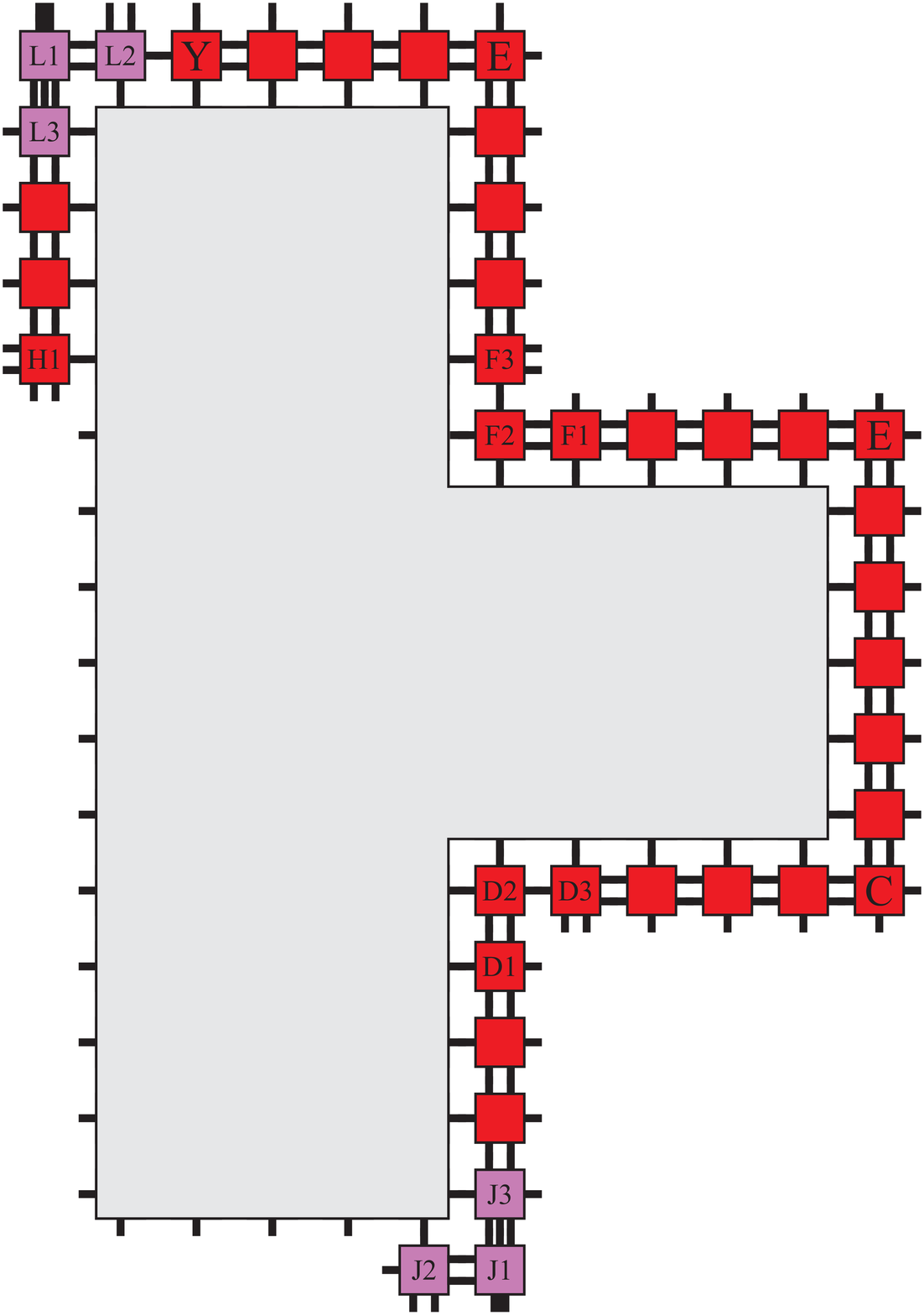}}\\
    \subfloat[][The order in which all of the DNA border tiles in \subref{fig:border_assembly_bad_2} eventually dissociate.]{
        \label{fig:border_disassembly_sequence}
        \includegraphics[width=2.0in]{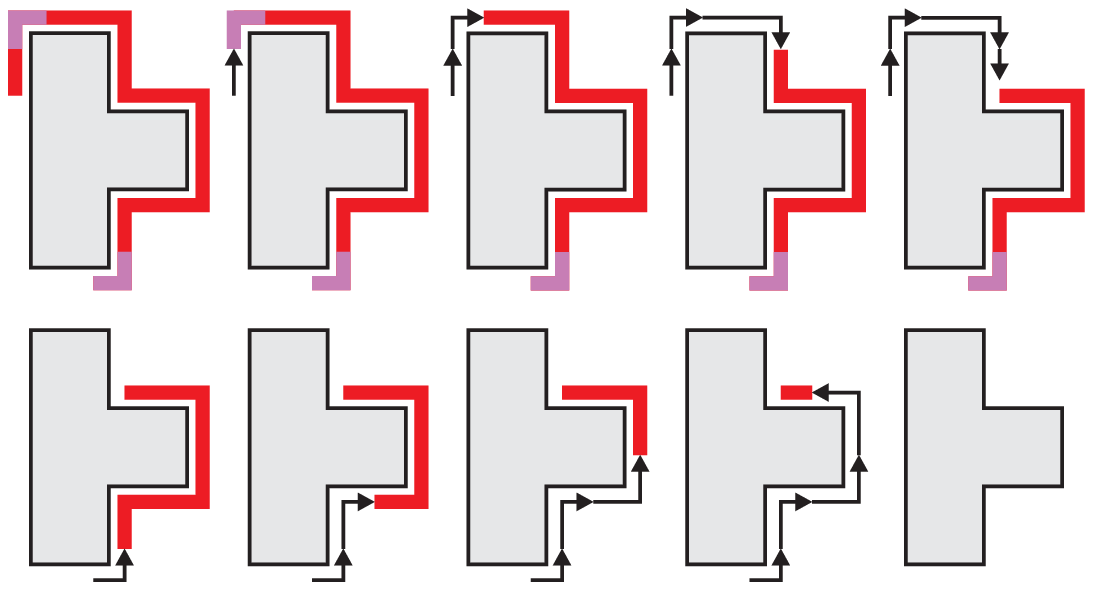}}
    \caption{\label{fig:border_assembly_bad} The border assembly phase in which the input shape is not the same as the target shape. }%
\end{figure}
\end{proof}

The number of unique tile types required for the unpacking phase is $O\left(\frac{\left| \pi_X \right|}{\log \left| \pi_X \right|}\right)$, while the number of tile types required for the remaining phases of the construction is a constant.  Hence, the identification complexity of $X$ with respect to $\mathcal{C}$ is $O\left( \frac{\left| \pi_X \right|}{\log \left| \pi_X \right|} \right)$.

Note that the restriction that $X$ be $x$-monotone is not necessary in the sense that the east and west sides of the input shape can have features similar to those of the north and south sides if the construction is extended in the obvious way. However, to make the class of identifiable shapes more straightforward to define and the construction easier to explain, we have presented the construction with that constraint.

\subsection{Discussion of Theorem~\ref{non_planar_more_shapes_two_stages}}
\label{two_dissolve_details}
\begin{figure}[htp]
    \centering
    \includegraphics[width=4.0in]{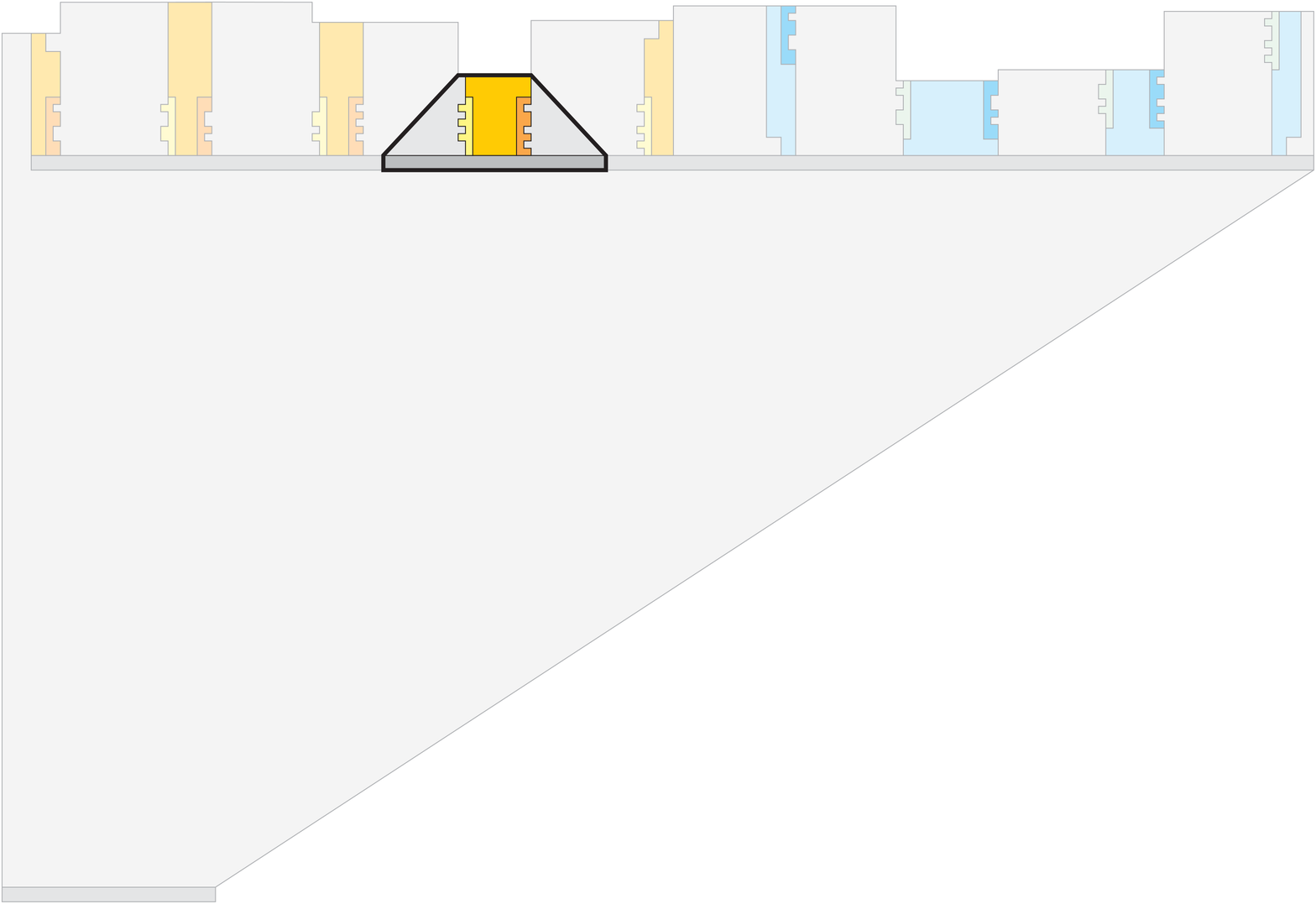}
    \caption{\label{fig:non_planar_two_dissolve_overview_2} A compact description of all of the rectangle supertiles that make up the frame is unpacked by a Turing machine. Then a constant size tile set is used to convert each description into a rectangle supertile with the appropriate connection interface.}
\end{figure}

\begin{figure}[htp]
\centering
        \includegraphics[width=\textwidth]{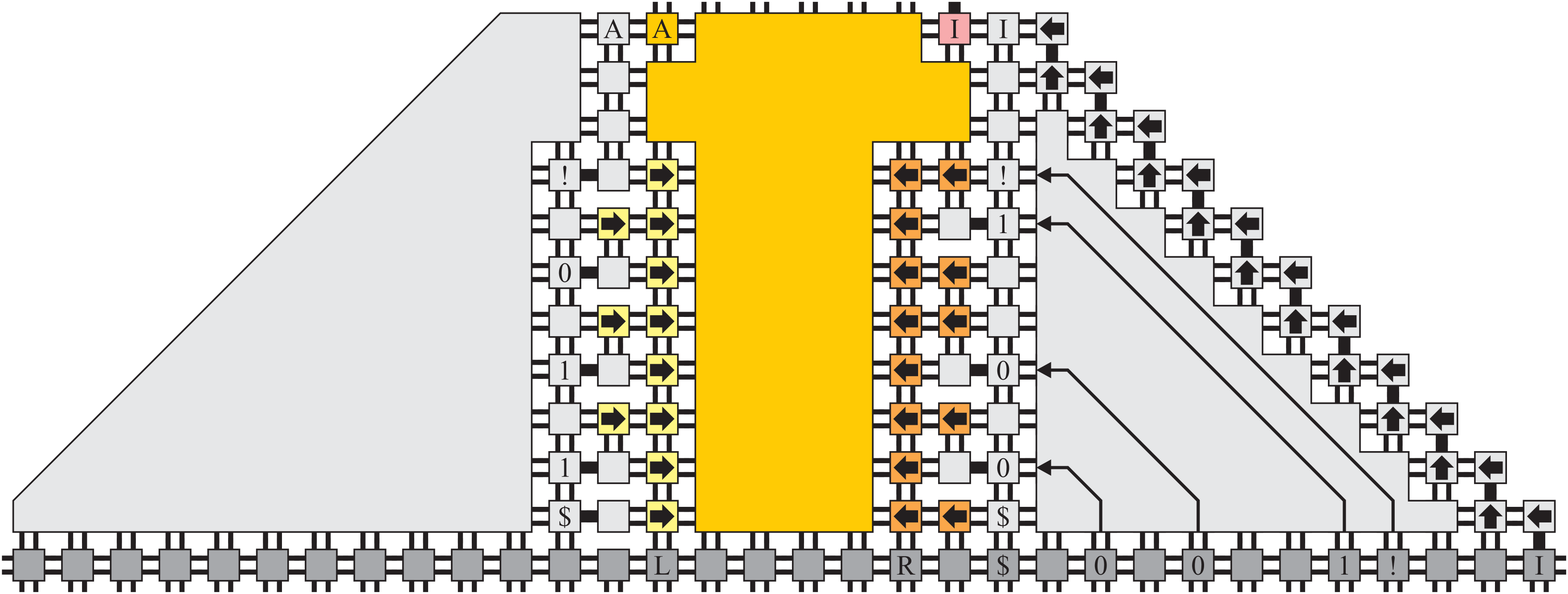}
        \caption{\label{fig:rectangle_supertile_detail} A closer look at the outlined portion of Figure~\ref{fig:non_planar_two_dissolve_overview_2}. Here we use a tile set to convert the algorithmic description of this particular rectangle supertile into an assembly of RNA tiles (of the type dissolved only by the second RNAse enzyme to be added). The arrows represent a standard self-assembly ``rotation'' scheme in which a horizontal sequence of bits (one bit per tile) is converted into a vertical sequence of bits.}
\end{figure}

\begin{figure}[htp]
    \centering
    \includegraphics[width=1.5in]{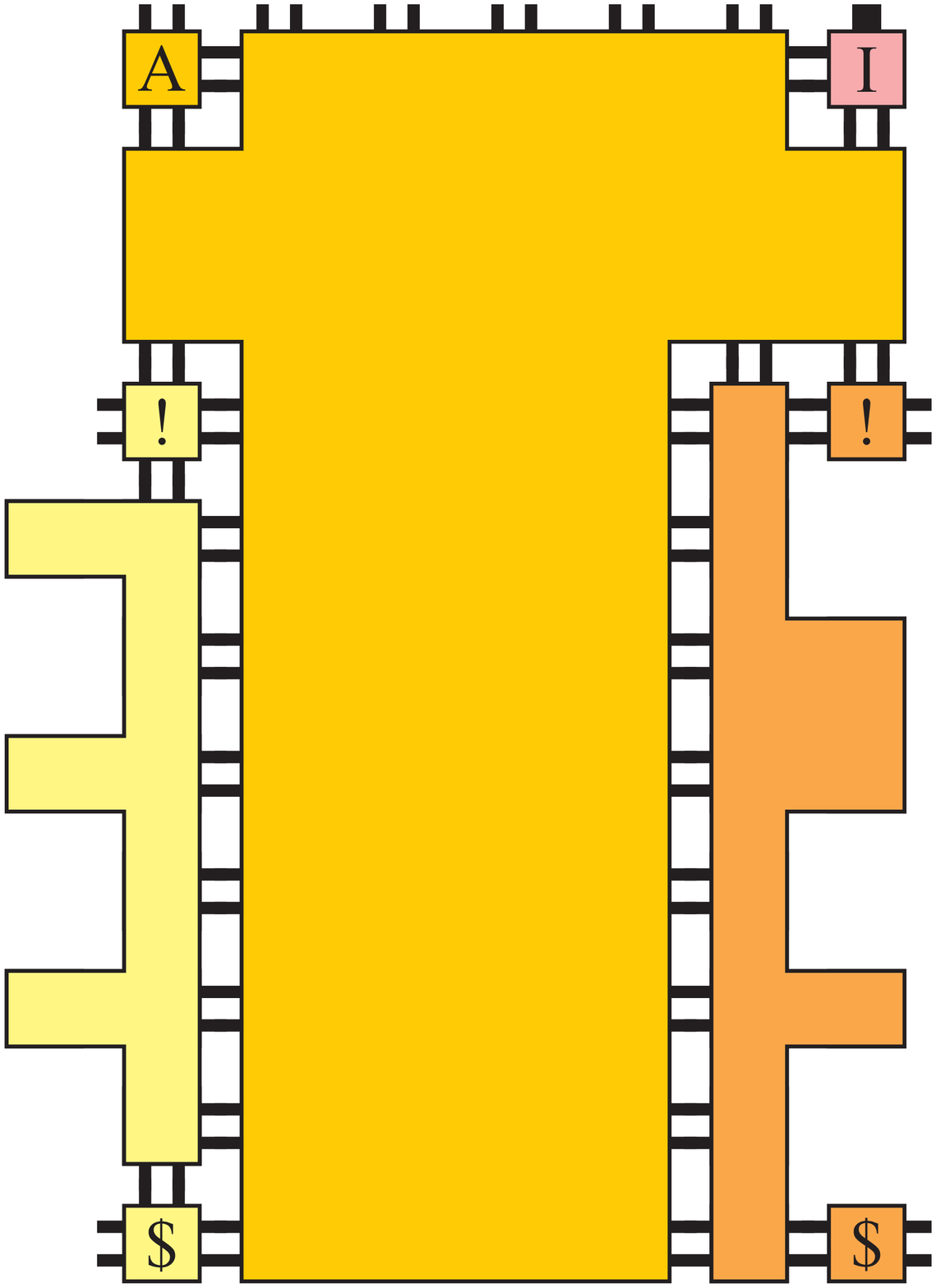}
    \caption{\label{fig:rectangle_supertile_alone} The rectangle supertile that is left after the first type of RNAse enzyme is added. Note that other rectangle supertiles attach via the ``!'' and ``?'' tiles with total strength $4$.}
\end{figure}

Prior to the first dissolve stage, we assemble pieces which will themselves eventually combine in a two-handed fashion to form the frame that accepts $X$ using a collection of rectangular supertiles of arbitrary dimension (we use a Turing machine to compute the dimensions of these rectangles using an algorithmic description of $X$). See Figure~\ref{fig:non_planar_two_overview}\subref{fig:non_planar_two_dissolve_overview_1} for an intuitive depiction of this process. After the the first type of RNAse enzyme is added, all of the rectangles assemble in a two-handed fashion and connect via ``binary teeth'' interfaces along their left and right sides (see \cite{DDFIRSS07,Replication10,RNAPods10} for examples of this well-known two-handed assembly technique). Note that these connection interfaces can be made arbitrarily long depending on how ``jagged'' the north- and south-facing features of $Y$ are. Once the frame assembles (see Figure~\ref{fig:non_planar_two_overview}\subref{fig:non_planar_two_dissolve_final}), a fully-connected border of DNA tiles will attach to $Y$ if and only if $Y = X$ in exactly the same fashion as it did in the construction for Theorem~\ref{non_planar_more_shapes}. After all of the border tiles attach, the second type of RNAse enzyme is added, dissolving the frame and completing the construction.

\begin{wrapfigure}{l}{1.25in}
    \vspace{-20pt}
\centering
    \includegraphics[width=1.0in]{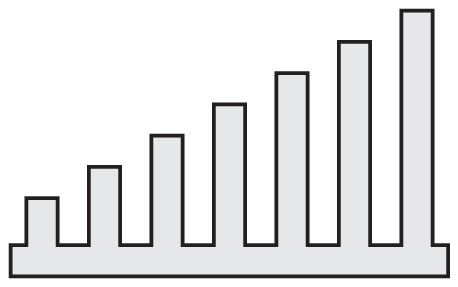} \caption{\label{fig:comb_example} A ``worst-case'' example target shape $X$. Note that we cannot re-use any rectangular supertiles that make up the frame that accepts $X$. }
    \vspace{-20pt}
\end{wrapfigure}

It is worthy of note that, in the construction discussed above, after the first type of RNAse enzyme is added, the frame that will eventually accept $X$ might be very large depending on how ``non-uniformly jagged'' the features of $Y$ are. This is because the more jagged $Y$is, the longer the rectangular supertiles that assemble into the frame that accepts $X$ must be in order to accommodate a larger number of unique connection interfaces. However, if many of the features of $X$ are ``similar,'' then connection interface patterns can be re-used thus eliminating the need for larger rectangular supertiles. An example of a kind of ``worst-case'' target shape $X$ might be a comb-like structure with a very long ``base'' and with each ``tooth'' slightly taller than the tooth to its left (see Figure~\ref{fig:comb_example}).

Similar to the previous construction, the number of unique tile types required for the unpacking phase is $O\left(\frac{\left| \pi_X \right|}{\log \left| \pi_X \right|}\right)$, while the number of tile types required for the remaining phases of the construction is a constant.  Hence, the identification complexity of $X$ with respect to $\mathcal{C}$ is $O\left( \frac{\left| \pi_X \right|}{\log \left| \pi_X \right|} \right)$.  Also similar to the previous construction, the restriction that $X$ be $x$-monotone isn't entirely necessary, and with the obvious modifications to the construction, many shapes with features on their east and west sides can also be identified.